\documentclass[journal]{IEEEtran}

\usepackage[dvips]{pstcol}
\usepackage{graphicx}
\usepackage{amsmath}

\usepackage{latexsym}
\usepackage{amsfonts}
\usepackage{amsbsy}
\usepackage{amssymb}
\usepackage{times}
\usepackage{enumerate}

\usepackage{subeqnarray}
\usepackage{cases}
\usepackage{stfloats}
\usepackage{cases}
\usepackage{subeqnarray}
\usepackage{amsmath}
\usepackage{multicol}
\usepackage{multirow}
\usepackage{bigstrut}
\usepackage{cite}
\usepackage{epstopdf}
\begin{document}
\title{Robust Transceiver Design for MISO Interference Channel with Energy Harvesting}

\author{Ming-Min Zhao, Yunlong Cai, Qingjiang Shi,  Benoit Champagne, and Min-Jian Zhao
\thanks{
Ming-Min Zhao, Yunlong Cai and Min-Jian Zhao are with the Department of Information Science and Electronic Engineering, Zhejiang University, Hangzhou 310027, China (e-mail: zmmblack@zju.edu.cn; ylcai@zju.edu.cn; mjzhao@zju.edu.cn).

Q. Shi is with the School of Info. Sci. \& Tech., Zhejiang Sci-Tech University, Hangzhou 310018, China. He is also with The State Key Laboratory of Integrated Services Networks, Xidian University (email: qing.j.shi@gmail.com).

B. Champagne is with the Department of Electrical and Computer Engineering, McGill University, Montreal, QC H3A 0E9, Canada (benoit.champagne@mcgill.ca).

This work was supported in part by the National Natural Science Foundation of China under Grant  61471319, Zhejiang Provincial Natural Science Foundation of China under Grant LY14F010013, the Fundamental Research Funds for the Central Universities, and the National High Technology Research and Development Program (863 Program) of China under Grant 2014AA01A707.
}
}

\maketitle
\begin{abstract}
In this paper, we consider
multiuser multiple-input single-output (MISO) interference channel where the received signal is divided into two parts for information decoding and energy harvesting (EH), respectively. The transmit beamforming vectors and receive power splitting (PS) ratios are jointly designed in order to minimize the total transmission power subject to both signal-to-interference-plus-noise ratio (SINR) and EH constraints. Most joint beamforming and power splitting (JBPS) designs assume that perfect channel state information (CSI) is available; however CSI errors are inevitable in practice. To overcome this limitation, we study the robust JBPS design problem assuming a norm-bounded error (NBE) model for the CSI. Three different solution approaches are proposed for the robust JBPS problem, each one leading to a different computational algorithm. Firstly, an efficient semidefinite relaxation (SDR)-based approach is presented to solve the highly non-convex JBPS problem, where the latter can be formulated as a semidefinite programming (SDP) problem. A rank-one recovery method is provided to recover a robust feasible solution to the original problem. Secondly, based on second order cone programming (SOCP) relaxation, we propose a low complexity approach with the aid of a closed-form robust solution recovery method.
Thirdly, a new iterative method is also provided which can achieve near-optimal performance when the SDR-based algorithm results in a higher-rank solution. We prove that this iterative algorithm monotonically converges to a Karush-Kuhn-Tucker (KKT) solution of the robust JBPS problem. Finally, simulation results are presented to validate the robustness and efficiency of the proposed algorithms.
\end{abstract}


\IEEEpeerreviewmaketitle

\section{Introduction}
Recently, energy harvesting (EH) from the environment has attracted considerable interest since it offers a promising solution to provide cost-effective and perpetual power supplies for wireless networks \cite{varshney2008transporting, grover2010shannon, OntheTransferofInformation, Twowaycommunication, MIMOBroadcastingfor, RobustBeamformingforWireless, MultiuserMISOBeamforming, ADynamicPowerSplittingApproach, ArchitectureDesign, EnergyharvestinginanOSTBC, Li2014Beamforming, Chen2015Distributed, JointTransmitBeamforming, Two-UserMIMOInterferenceChannel, K-UserMIMOInterferenceChannel, BeamformingforMISO, JointBeamformingAnd, Mguyen2015Opportunistic}.
The unified study of simultaneous wireless information and power transfer (SWIPT) has therefore drawn significant attention lately, as it opens new challenges and possibilities in the analysis and design of transmission schemes and protocols.

The fundamental concept of SWIPT was first proposed in \cite{varshney2008transporting}, which characterizes the rate-energy (R-E) tradeoff in a discrete memoryless channel. The study of R-E tradeoff was later extended to frequency selective channels \cite{grover2010shannon}, multiple access and multi-hop channels \cite{OntheTransferofInformation}, and two-way channels \cite{Twowaycommunication}. However, the above works all assume that the receiver can decode information and harvest energy from the same signal, which is impossible with existing technology. In \cite{MIMOBroadcastingfor}, the authors proposed a practical receiver structure for the first time, and considered the R-E region and optimal transmission scheme of a MIMO broadcasting channel. Two practical signal separation schemes, namely time switching (TS) and power splitting (PS) were also considered.
For the TS scheme, the transmitter divides the transmission block into two orthogonal time slots, one for transferring power and the other for transmitting data. For the PS scheme, the received signal is split with an adjustable PS ratio, where the stream with power ratio $0 \le \rho  \le 1$ is used for information decoding (ID) and the other stream with power ratio $1 - \rho $ is used for EH.

The works in \cite{RobustBeamformingforWireless} and \cite{MultiuserMISOBeamforming} considered beamforming designs with separate information/energy receivers. Specially, \cite{RobustBeamformingforWireless} studied the robust beamforming problem for the multi-antenna wireless broadcasting system with SWIPT, under the assumption of imperfect channel state information (CSI) at the transmitter. In \cite{MultiuserMISOBeamforming} the authors investigated the optimal beamforming strategy to maximize the weighted sum-power transferred to all EH receivers subject to given minimum signal-to-interference-plus-noise ratio (SINR) constraints at different ID receivers.
The work in \cite{ADynamicPowerSplittingApproach} derived the optimal power splitting rule at the receiver to achieve various trade-offs between the maximum ergodic capacity for information transfer and the maximum average EH for power transfer. In \cite{ ArchitectureDesign}, various practical receiver architectures for SWIPT were investigated, where a new integrated information and energy receiver design was proposed. The use of SWIPT for relay systems was considered in \cite{EnergyharvestinginanOSTBC, Li2014Beamforming, Chen2015Distributed}.

The work \cite{JointTransmitBeamforming} considers a multiuser MISO downlink system with SWIPT, where the total transmission power at the base station (BS) is minimized subject to given SINR and EH constraints. An optimal solution is proposed based on semidefinite relaxation (SDR) along with low-complexity suboptimal designs.
Some recent studies have focused on SWIPT in the context of multi-antenna interference channels \cite{Two-UserMIMOInterferenceChannel, K-UserMIMOInterferenceChannel, BeamformingforMISO, JointBeamformingAnd}. The work \cite{Two-UserMIMOInterferenceChannel} investigates optimal transmission strategies and mode scheduling methods for a two-user MIMO interference channel with EH, while \cite{K-UserMIMOInterferenceChannel} extends this study to the $K$-user MIMO interference channel. A number of recent studies focus on the PS approach and consider the joint beamforming and power splitting (JBPS) design problem in MISO interference channel, where the downlink receivers are characterized by both SINR and EH constraints. Compared with the conventional beamforming design, JBPS design is much more challenging due to the coupling between the beamforming vectors and PS ratios. Hence, the corresponding research area has become quite active and several algorithms have been recently proposed to address this problem. In \cite{BeamformingforMISO} and \cite{JointBeamformingAnd}, the JBPS design is studied for a $K$-user MISO interference channel with the same design criterion as that in \cite{JointTransmitBeamforming}. Specially, the work \cite{BeamformingforMISO} uses SDR to tackle the non-convex JBPS problem and prove that the SDR is tight when $K=2$ or $3$. Also, various suboptimal but low complexity solutions based on fixed beamforming schemes and a hybrid scheme are provided. In \cite{JointBeamformingAnd}, the JBPS problem is reformulated as a SOCP problem based on an alternative method named SOCP relaxation and two sufficient conditions are given under which the relaxation is tight. Also, a primal-decomposition based distributed algorithm is proposed for the JBPS problem.

In these existing works on JBPS, the CSI is assumed to be perfectly known. In practice however, the CSI is prone to errors owing to various factors, which may limit the system performance drastically. Hence, it is of interest to develop JBPS designs that are robust to CSI errors. In this paper, assuming norm-bounded CSI error (NBE) models for the CSI, we propose three new computational algorithms for solving the robust JBPS design problem in a $K$-user MISO interference channel with multi-antenna transmit beamformers and single-antenna PS receivers. In the first approach, we show that the robust JBPS problem can be relaxed as a SDP problem based on SDR. A rank-one recovery method is provided to recover a robust feasible solution to the original problem if a high-rank solution is returned. In the second approach, we propose to formulate the original problem as a SOCP problem based on SOCP relaxation and the Cauchy-Schwarz inequality for the purpose of complexity reduction. Since the solution to the SOCP problem is not necessarily robust, a closed-form robust solution recovery method is provided. Finally, as our third approach, we propose a new iteration algorithm based on the concave-convex procedure (CCCP) which can provide near-optimal performance when higher-rank solutions are returned by the SDR-based algorithm. The convergence of this iterative algorithm is studied in detail and we prove that any limit point of the iterative algorithm is a Karush-Kuhn-Tucker (KKT) solution to the robust JBPS problem. Finally, simulation results are presented to validate the robustness and efficiency of the proposed algorithms. In particular, we show that the proposed robust algorithms can provide near-optimal performance.

The reminder of this paper is organized as follows. We present the system model of the $K$-user MISO interference channel, the channel error model and the robust JBPS problem in Section \ref{syste-mmodel}. In Section \ref{Section_SDP}, the robust JBPS design approaches based on SDR and SOCP relaxation are developed along with their rank-one recovery methods and final computational algorithms. In Section \ref{Section_iterative}, we develop the CCCP-based iterative design algorithm for the JBPS problem and discuss its convergence and initialization. A detailed complexity analysis of the proposed algorithms is provided in Section \ref{section:complexity}. Finally, in Section \ref{section:simulation} computer simulations are used to compare the robust performance of the proposed JBPS designs. Conclusions are drawn in Section \ref{section:conclusion}.

\emph{Notations:} Scalars, vectors and matrices are respectively denoted by lower case, boldface lower case and boldface upper case letters. For a square matrix $\bf{A}$, $\textrm{Tr} (\bf{A})$, $\textrm{rank}\bf(A)$, ${{\bf{A}}^T}$, $\textrm{conj}\bf(A)$, ${{\bf{A}}^H}$, ${{\bf{A}}^{ - 1}}$ and ${{\bf{A}}^\dag }$ denote its trace, rank, transpose, conjugate,  conjugate transpose, inverse and pseudo-inverse, respectively, while ${\bf{A}} \succeq {\bf{0}}$ means that $\bf{A}$ is a positive semidefinite matrix. The operator $\textrm{diag} \{{\bf{S}}_1, \ldots , {\bf{S}}_M\}$ denotes a block-diagonal square matrix with ${\bf{S}}_1, \ldots , {\bf{S}}_M$ denoting the diagonal square matrices. The operator $\textrm{vec}(\cdot)$ stacks the elements of a matrix in one long column vector, ${\textrm{invp}}( x )$ denotes the inverse of the positive portion, i.e. $\frac{1}{{\max ( {x,0} )}}$. $\|\cdot\|$ denotes the Euclidean norm of a complex vector and $|\cdot|$ denotes the absolute value of a complex scalar. Finally, ${\mathbb{C}^{m \times n}}\;({\mathbb{R}^{m \times n}})$ denotes the space of ${m \times n}$ complex (real) matrices, and $\mathbb{R}_+\;(\mathbb{R}_-)$ denotes the space of positive (negative) real numbers.
\section{System Model And Problem Formulation} \label{syste-mmodel}
\subsection{Proposed System Model}
\label{Section2:system}
\begin{figure}[hbtp]
  \centering
  \includegraphics[width = 0.42\textwidth]{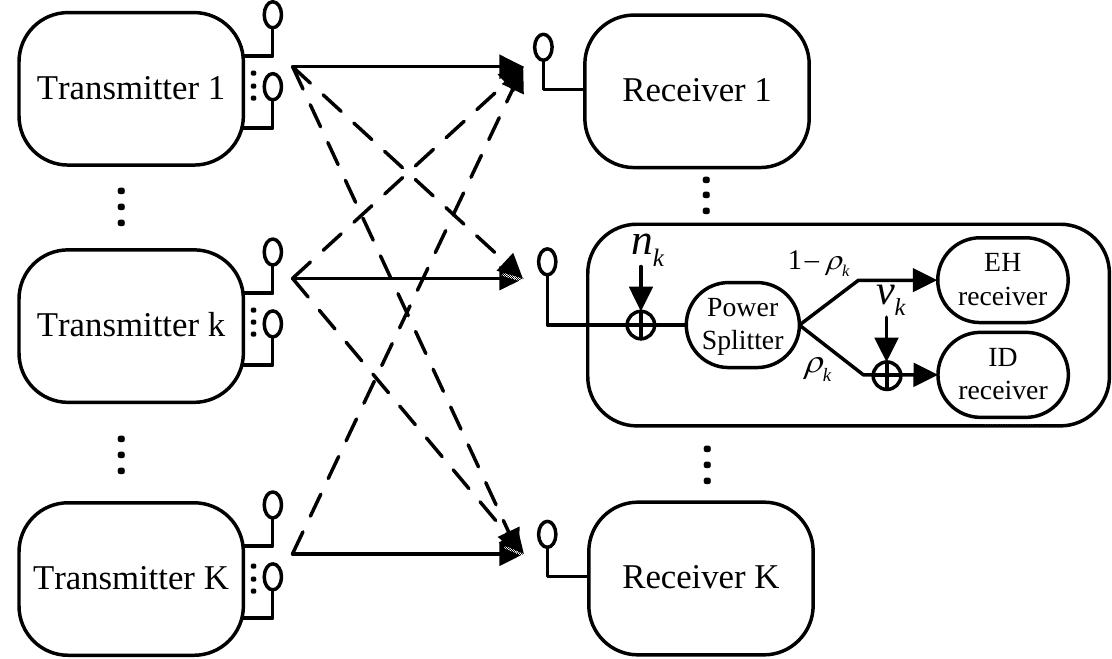}
  \caption{The $K$-user MISO interference channel. Each receiver splits the received energy in two parts for ID and EH, respectively.}\label{systemmodel}
\end{figure}

We consider the $K$-user MISO interference channel where each transmitter, indexed by $k \in \mathcal{K} \triangleq \{1,\ldots,K\}$, is equipped with $N_k$ antennas and each receiver is equipped with a single antenna. The $K$ transmitters are assumed to operate over a common frequency band and each communicates with its corresponding receiver using transmit beamforming. Different from conventional interference channels, we here consider PS-based receivers. The received signal at each receiver is split into two separate signal streams with different power levels, one sent to the EH receiver and the other to the ID receiver \cite{MIMOBroadcastingfor}. The system model is illustrated in Fig. \ref{systemmodel}. We assume that transmitter $k$ sends its signal $s_k$ to its intended receiver through beamforming vector ${{\bf{f}}_k} \in {\mathbb{C} ^{{N_k} \times 1}}$, and that the ${s_k}$ are statistically independent with zero mean and $E\{ {{{\left| {{s_k}} \right|}^2}} \} = 1$ for all $k \in \mathcal{K}$.
Under these conditions, the available baseband signal at receiver $k$ before PS is ideally given by\\[-3mm]
\begin{equation}
{r_k} = \underbrace {{\bf{h}}_{kk}^H{{\bf{f}}_k}{s_k}}_{desired\;signal} + \underbrace {\sum\limits_{j = 1,j \ne k}^K {{\bf{h}}_{kj}^H{{\bf{f}}_j}{s_j}} }_{interference} + {n_k},
\end{equation}
\\[-3mm]
where ${{\bf{h}}_{kj}} \in {\mathbb{C}^{{N_k} \times 1}}$ denotes the channel vector between transmitter $j$ and receiver $k$, and ${n_k} \in \mathbb{C}$ is the additive white Gaussian noise (AWGN) introduced by the receive antenna, which is assumed to have zero mean and variance $\sigma _k^2$.

Each receiver splits its received signal to the information decoder and the energy harvester by means of a power splitter. Let ${\rho _k}\left( {0 \le {\rho _k} \le 1} \right)$ denote the power splitting (PS) ratio for receiver $k$, which means that a portion ${\rho _k}$ of the signal power is used for signal detection while the remaining portion $1 - {\rho _k}$ is diverted to an energy harvester. Thus, the available signal for ID at receiver $k$ can be expressed as
\begin{equation}
r_k^{\textrm{ID}} = \sqrt {{\rho _k}} \left( {{\bf{h}}_{kk}^H{{\bf{f}}_k}{s_k} + \sum\limits_{j = 1,j \ne k}^K {{\bf{h}}_{kj}^H{{\bf{f}}_j}{s_j}}  + {n_k}} \right) + {v_k},
\end{equation}
where ${v_k}$ is the additional AWGN circuit noise with zero mean and variance $\omega_k^2$ due to phase offset and non-linearities during baseband conversion \cite{MIMOBroadcastingfor}. We assume that each receiver employs single-user detection by treating the cochannel interference as noise. Then, the SINR at receiver $k$ is given by \\[-3mm]
\begin{equation} \\[-2mm]
{\Gamma _k} = \frac{{{\rho _k}{{| {{\bf{h}}_{kk}^H{{\bf{f}}_k}} |}^2}}}{{{\rho _k}\left( {\sum\limits_{j = 1,j \ne k}^K {{{| {{\bf{h}}_{kj}^H{{\bf{f}}_j}} |}^2}}  + \sigma _k^2} \right) + \omega _k^2}}.
\end{equation}
Besides, the total harvested energy that can be stored by receiver $k$ is equal to \\[-3mm]
\begin{equation} \\[-2mm]
P_k^{\textrm{EH}} = {\xi _k}\left( {1 - {\rho _k}} \right)\left( {\sum\limits_{j = 1}^K {{{| {{\bf{h}}_{kj}^H{{\bf{f}}_j}} |}^2}}  + \sigma _k^2} \right),
\end{equation}
where ${\xi _k} \in \left( {0,1} \right]$ denotes the energy conversion efficiency of the $k$th EH unit, which means that a portion ${\xi _k}$ of the received RF signal is used for EH.

 \vspace{-1.4em}

\subsection{Channel Error Model}
Because of many factors such as channel estimation errors, quantization errors, and feedback errors/delay, it is not possible in practice to obtain perfect CSI at both the transmitters and receivers. Let ${\widehat {\bf{h}}_{kj}} \in {\mathbb{C}^{{N_k} \times 1}}$ denotes the estimated channel vector between transmitter $j$ and receiver $k$. Then, the actual CSI can be expressed as \\[-3mm]
\begin{equation} \\[-1mm]
{{\bf{h}}_{kj}} = {\widehat {\bf{h}}_{kj}} + {{\bf{e}}_{kj}},\quad j,k \in \mathcal{K},
\end{equation}
where ${{\bf{e}}_{kj}}$ denotes the CSI error vector. To model the CSI errors, the well-known norm-bounded error (NBE) model \cite{Relayprecoder} is adopted, where we assume that the channel estimation error ${{\bf{e}}_{kj}}$ is bounded in its Euclidean norm, that is \\[-3mm]
\begin{equation} \\[-1mm]
\| {{{\bf{e}}_{kj}}} \| \le {\eta _{kj}},{\mkern 1mu} {\mkern 1mu} \quad j,k \in \mathcal{K},
\end{equation}
where ${\eta _{kj}}$ is a known positive constant. Equivalently, ${{\bf{h}}_{kj}}$ belongs to the uncertainty set ${\Re _{kj}}$ defined as
\begin{equation} \label{channelerror}
{\Re _{kj}} = \{ {{\bf{h}}| {{\bf{h}} = {{\widehat {\bf{h}}}_{kj}} + {{\bf{e}}_{kj}}},\;\| {{{\bf{e}}_{kj}}} \| \le {\eta _{kj}}} \}.
\end{equation}

The shape and size of $\Re _{kj}$ model the kind of uncertainty in the estimated CSI, which is linked to the physical phenomenon producing the CSI errors. It should be emphasized that the actual errors $\{{\bf{e}}_{kj}\}$ are assumed to be unknown while the corresponding upper bounds $\{\eta _{kj}\}$ can be obtained using the preliminary knowledge of the type of imperfection and/or coarse knowledge of the channel type and its main characteristics \cite{Robustdownlinkpower}. This model is particularly suitable for systems where CSI is corrupted by quantization \cite{RobustPowerallocation}.

\subsection{Optimization Problem}
We assume that the ID and EH units of each receiver are characterized by certain quality of service (QoS) and EH constraints. The QoS constraints require that the SINR of receiver $k$ should be no smaller than a given positive target ${\gamma _k}$. In the meantime, the EH constraints call for the harvested energy of receiver $k$ to be no smaller than a positive threshold ${\psi _k}$. In this work, we focus on robust JBPS design under such constraints. Specifically, minimization of the total transmission power subject to the above two types of constraints, in the presence of imperfect CSI with NBE model, can be formulated as the following constrained optimization problem:
\begin{equation} \label{robustproblem}
\begin{array}{*{20}{l}}
{\begin{array}{*{20}{c}}
{\mathop {\min }\limits_{\left\{ {{{\bf{f}}_k},\;{\rho _k}} \right\}} }&{\sum\limits_{k = 1}^K {{{\| {{{\bf{f}}_k}} \|}^2}} }
\end{array}}\\
{\begin{array}{*{20}{c}}
{{\textrm{s}}.{\textrm{t}}.}&{\begin{array}{*{20}{l}}
{\frac{{{\rho _k}{{| {{\bf{h}}_{kk}^H{{\bf{f}}_k}} |}^2}}}{{{\rho _k}\left( {\sum\limits_{j = 1,j \ne k}^K {{{| {{\bf{h}}_{kj}^H{{\bf{f}}_j}} |}^2}}  + \sigma _k^2} \right) + \omega _k^2}} \ge {\gamma _k},}\\
{{\xi _k}\left( {1 - {\rho _k}} \right)\left( {\sum\limits_{j = 1}^K {{{| {{\bf{h}}_{kj}^H{{\bf{f}}_j}} |}^2}}  + \sigma _k^2} \right) \ge {\psi _k},}\\
{0 \le {\rho _k} \le 1,\;\| {\bf{e}}_{kj} \|^2 \le \eta _{kj}^2,\;\forall j,k \in \mathcal{K}.}
\end{array}}
\end{array}}
\end{array}
\end{equation}


Different from \cite{BeamformingforMISO,JointBeamformingAnd}, where it is assumed that the perfect knowledge of CSI is available, the goal of our work is to investigate the robust JBPS design, i.e. to guarantee that the SINR targets ${\gamma _k}$ and EH requirement ${\psi _k}$ are satisfied for all possible CSI errors. As compared to the non-robust design in \cite{BeamformingforMISO,JointBeamformingAnd}, the above robust JBPS design can provide guaranteed QoS and harvested energy for each receiver for all possible channel realizations in the bounded uncertainty regions ${\Re _{kj}}$ in (\ref{channelerror}). Solving the robust design problem (\ref{robustproblem}), however, is more challenging because there is an infinite number of constraints (due to the NBE model) and each of the SINR (EH) constraint is not convex. Both these properties make the problem (\ref{robustproblem}) very difficult to address.

Similar to Lemma 3.1 \& Lemma 3.2 in \cite{JointTransmitBeamforming}, it can be shown that the feasibility of (\ref{robustproblem}) is independent of the EH constraints and PS ratios. That is, problem (\ref{robustproblem}) is feasible as long as the following problem is feasible.
\begin{equation} \label{robustproblem_fea}
\begin{array}{*{20}{l}}
{\textrm{find} \quad \{ {{{\bf{f}}_k}} \}}\\
{\begin{array}{*{20}{c}}
{{\textrm{s}}.{\textrm{t}}.}&\begin{array}{l}
\frac{{{{| {{\bf{h}}_{kk}^H{{\bf{f}}_k}} |}^2}}}{{\sum\limits_{j = 1,j \ne k}^K {{{| {{\bf{h}}_{kj}^H{{\bf{f}}_j}} |}^2}}  + \sigma _k^2 + \omega _k^2}} \ge {\gamma _k},\\
\| {\bf{e}}_{kj} \|^2 \le \eta _{kj}^2,\;\forall j,k \in \mathcal{K}.
\end{array}
\end{array}}
\end{array}
\end{equation}
This property provides an easier way to check the feasibility of (\ref{robustproblem}). In the rest of this paper, we assume (\ref{robustproblem_fea}) is feasible.

\section{Proposed Robust Design Based on Relaxation} \label{Section_SDP}
In this section, we propose two efficient approaches to solve the robust JBPS design problem (\ref{robustproblem}). In the first design approach, the celebrated SDR technique is applied to convert the semi-infinite constraints into linear matrix inequalities by means of the S-Procedure \cite{ConvexOptimization, boyd1994linear}.
In general, the relaxed problem is not guaranteed to have a rank-one solution \cite{BeamformingforMISO, Distributedrobustmulti, OptimizationofMIMOrelays}, we thus provide a good heuristic solution to recover a feasible solution to (\ref{robustproblem}). In the second design approach, we formulate problem (\ref{robustproblem}) into a SOCP problem based on SOCP relaxation. The resulting SOCP problem has very low computational complexity and shows great potential for applications with large antenna arrays and large number of transmit-receive pairs.

\subsection{SDP with Rank Relaxation} \label{section:SDP}
According to \cite{Semidefiniterelaxation}, we introduce a new optimization variable ${{\bf{F}}_k} \buildrel \Delta \over = {{\bf{f}}_k}{\bf{f}}_k^H,\;\forall k \in \mathcal{K}$ and rewrite problem (\ref{robustproblem}) as follows \\[-2mm]
\begin{equation} \label{SDRproblem} \\[-1mm]
\begin{array}{*{20}{l}}
{\begin{array}{*{20}{c}}
{\mathop {\min }\limits_{\{ {{{\bf{F}}_k},\;{\rho _k}} \}} }&{\sum\limits_{k = 1}^K {{\textrm{Tr}}( {{{\bf{F}}_k}} )} }
\end{array}}\\
{\begin{array}{*{20}{l}}
{{\textrm{s}}.{\textrm{t}}.\;\;\frac{1}{{{\gamma _k}}}{\bf{h}}_{kk}^H{{\bf{F}}_k}{\bf{h}}_{kk}^{} - \sum\limits_{j \ne k}^K {{\bf{h}}_{kj}^H{{\bf{F}}_j}{\bf{h}}_{kj}^{}}  \ge \sigma _k^2 + \frac{{\omega _k^2}}{{{\rho _k}}},}\\
{\begin{array}{*{20}{l}}
& {\sum\limits_{j = 1}^K {{\bf{h}}_{kj}^H{{\bf{F}}_j}{\bf{h}}_{kj}^{}}  \ge \frac{{{\psi _k}}}{{{\xi _k}( {1 - {\rho _k}} )}} - \sigma _k^2,} \\
& {0 \le {\rho _k} \le 1,\;{{\bf{F}}_k} \succeq {\bf{0}},\;{\textrm{rank}}( {{{\bf{F}}_k}} ) = 1,}\\
& { \| {\bf{e}}_{kj} \|^2 \le \eta _{kj}^2,\;\forall j,k \in \mathcal{K}.}
\end{array}}
\end{array}}
\end{array}
\end{equation}
\\[-2mm]
This problem can be relaxed as a convex problem by dropping the non-convex rank-one constraint $\textrm{rank}( {{{\bf{F}}_k}} ) = 1$, since the objective function and the constraints are linear in ${{\bf{F}}_k}$ and $\frac{1}{{{\rho _k}}},\frac{1}{{1 - {\rho _k}}}$ are both convex function of $\rho _k$. It is worth noting that the relaxation is not optimum and postprocessing of the relaxed problem will be discussed in detail in Subsection \ref{sec_rankrecovery}. However, this problem is still computationally intractable because it involves an infinite number of constraints.
By applying the S-Procedure, the infinitely many constraints can be reformulated into finite convex constraints.

We first observe that each term in the SINR and EH constraints containing ${{\bf{F}}_k}$ in problem (\ref{SDRproblem}) involves independent CSI errors. Hence, we introduce two auxiliary variables
\begin{equation} \\[-2mm]
{p_{kj}} = \mathop {\max }\limits_{\forall {\bf{e}}_{kj}^H{\bf{e}}_{kj}^{} \le {\eta _{kj}^2}} \;{\bf{h}}_{kj}^H{{\bf{F}}_j}{\bf{h}}_{kj}^{},\;k \in \mathcal{K},\;j \ne k,
\end{equation}
\begin{equation}\\[-2mm]
{q_{kj}} = \mathop {\min }\limits_{\forall {\bf{e}}_{kj}^H{\bf{e}}_{kj}^{} \le {\eta _{kj}^2}} \;{\bf{h}}_{kj}^H{{\bf{F}}_j}{\bf{h}}_{kj}^{},\;k \in \mathcal{K},\;j \ne k,
\end{equation}
where ${p_{kj}}$ is the maximum (worst-case) cochannel interference power from transmitter $j$ to receiver $k$ and ${q_{kj}}$ denotes the minimum (worst-case) power available for EH from the transmitter $j$ to receiver $k$. Then, with the help of these two variables, the SINR constraints in problem (\ref{SDRproblem}) can be equivalently rewritten as \\[-3mm]
\begin{flalign} \label{SINRcons1}
\begin{split}
& \;\;\; \frac{1}{{{\gamma _k}}}( {\widehat {\bf{h}}_{kk}^H + {\bf{e}}_{kk}^H} ){{\bf{F}}_k}( {\widehat {\bf{h}}_{kk}^{} + {\bf{e}}_{kk}^{}} )\\
& \;\;\; \ge \sum\limits_{j = 1,j \ne k}^K {{p_{kj}}}  + \sigma _k^2 + \frac{{\omega _k^2}}{{{\rho _k}}},\; \| {\bf{e}}_{kk} \|^2 \le \eta _{kk}^2,\forall k \in \mathcal{K},
\end{split} &
\end{flalign}
\vspace{-1em}
\begin{flalign} \label{SINRcons2}
\begin{split}
& \;\;\; ({\widehat {\bf{h}}_{kj}^H + {\bf{e}}_{kj}^H} ){{\bf{F}}_j}( {\widehat {\bf{h}}_{kj}^{} + {\bf{e}}_{kj}^{}} ) \le {p_{kj}},\;\\
& \;\;\; \| {\bf{e}}_{kj} \|^2 \le \eta _{kj}^2,\;\forall k \in \mathcal{K},j \ne k.
\end{split} &
\end{flalign}

Similarly, the EH constraints in problem (\ref{SDRproblem}) can be expressed as
\begin{flalign} \label{EHcons1}
\begin{split}
&\;\;\;( {\widehat {\bf{h}}_{kk}^H + {\bf{e}}_{kk}^H} ){{\bf{F}}_k}( {\widehat {\bf{h}}_{kk}^{} + {\bf{e}}_{kk}^{}} ) + \sum\limits_{j = 1,j \ne k}^K {{q_{kj}}} \\ &\;\;\;\;\ge \frac{{{\psi _k}}}{{{\xi _k}( {1 - {\rho _k}} )}} - \sigma _k^2,\; \| {\bf{e}}_{kk} \|^2 \le \eta _{kk}^2,\;\forall k \in \mathcal{K},
\end{split} &
\end{flalign}
\begin{equation}\label{EHcons2}
\begin{array}{l}
(\widehat {\bf{h}}_{kj}^H + {\bf{e}}_{kj}^H){{\bf{F}}_j}(\widehat {\bf{h}}_{kj}^{} + {\bf{e}}_{kj}^{}) \ge {q_{kj}},\\
\|{{\bf{e}}_{kj}}\|{^2} \le \eta _{kj}^2,\;\forall k \in {\mathcal K},j \ne k.
\end{array}
\end{equation}

By applying the S-Procedure, the constraints in (\ref{SINRcons1}) and (\ref{SINRcons2}) can be reformulated to finite convex constraints, which are equivalent to (\ref{SINRcons1_SDP}) and (\ref{SINRcons2_SDP}), shown at the top of the next page, where ${\alpha _k} = \frac{1}{{{\rho _k}}}$, and ${\lambda _{kj}},\;\forall j,k \in \mathcal{K}$ are slack variables. Similarly, we can recast (\ref{EHcons1}) and (\ref{EHcons2}) as (\ref{EHcons1_SDP}) and (\ref{EHcons2_SDP}) with  ${\beta _k} = \frac{1}{{1-{\rho _k}}}$, and slack variables ${\mu _{kj}},\;\forall j,k \in \mathcal{K}$. Then, problem (\ref{SDRproblem}) can be expressed as \\[-1mm]
\begin{equation} \label{robustSDP} \\[-1mm]
\begin{array}{*{20}{l}}
{\begin{array}{*{20}{c}}
{\mathop {\min }\limits_{\{ {{\bf{F}}_k},\; {\alpha _k},\; {\beta _k},\; {\lambda _{kj}},\; {\mu _{kj}},\},\;\{ {p_{kj}},\; {q_{kj}} \}_{j\neq k}} }&{\sum\limits_{k = 1}^K {{\textrm{Tr}}( {{{\bf{F}}_k}} )} }
\end{array}}\\
{\begin{array}{*{20}{c}}
{\textrm{s.t.}}\;\begin{array}{l}
(\ref{SINRcons1_SDP}), (\ref{SINRcons2_SDP}), (\ref{EHcons1_SDP})\; \text{and} \; (\ref{EHcons2_SDP}),\\
{\alpha _k} \ge 1,\;{\beta _k} \ge 1,\;{\textrm{invp}}( {{\alpha _k}} ) + {\textrm{invp}}( {{\beta _k}} ) \le 1,\\
p_{kj} \geq 0,\; q_{kj} \geq 0,\; j \neq k,\\
{{\bf{F}}_k} \succeq {\bf{0}},\;{\lambda _{kj}} \ge 0,\;{\mu _{kj}} \ge 0,\;\forall j,k \in \mathcal{K},
\end{array}
\end{array}}
\end{array}
\end{equation}
where constraints (\ref{SINRcons1_SDP})-(\ref{EHcons2_SDP}) appear at the top of the next page. In (\ref{robustSDP}), the set of constraints involving ${\textrm{invp}}(  \cdot  )$ must be satisfied with equality at optimality; otherwise the objective value can be further decreased by decreasing ${\alpha _k}^\prime$s. The above problem is a convex SDP problem which can be solved by an off-the-shelf solver \cite{UsingSeDuMi, Solvingsemidefinite}.

\begin{figure*}
\begin{equation} \label{SINRcons1_SDP}
{{\bf{U}}_k}( {{{\bf{F}}_k},{{\{ {{p_{kj}}} \}}_{j \ne k}},{\lambda _{kk}}}, {{\alpha _k}} ) \buildrel \Delta \over = \left[ {\begin{array}{*{20}{c}}
{\frac{1}{{{\gamma _k}}}{{\bf{F}}_k} + {\lambda _{kk}}{\bf{I}}}&{\frac{1}{{{\gamma _k}}}{{\bf{F}}_k}{{\widehat {\bf{h}}}_{kk}}}\\
{\frac{1}{{{\gamma _k}}}\widehat {\bf{h}}_{kk}^H{{\bf{F}}_k}}&{\frac{1}{{{\gamma _k}}}\widehat {\bf{h}}_{kk}^H{{\bf{F}}_k}{{\widehat {\bf{h}}}_{kk}} - \sum\limits_{j = 1,j \ne k}^K {{p_{kj}}}  - \sigma _k^2 - \omega _k^2{\alpha _k} - {\lambda _{kk}}\eta _{kk}^2}
\end{array}} \right]\succeq \bf{0}
\end{equation}
\begin{equation} \label{SINRcons2_SDP}
{{\bf{V}}_{kj}}( {{{\bf{F}}_j},{p_{kj}},{\lambda _{kj}}} ) \buildrel \Delta \over = \left[ {\begin{array}{*{20}{c}}
{ - {{\bf{F}}_j} + {\lambda _{kj}}{\bf{I}}}&{ - {{\bf{F}}_j}{{\widehat {\bf{h}}}_{kj}}}\\
{ - \widehat {\bf{h}}_{kj}^H{{\bf{F}}_j}}&{{p_{kj}} - \widehat {\bf{h}}_{kj}^H{{\bf{F}}_j}{{\widehat {\bf{h}}}_{kj}} - {\lambda _{kj}}\eta _{kj}^2}
\end{array}} \right] \succeq {\bf{0}},j \ne k
\end{equation}
\begin{equation} \label{EHcons1_SDP}
{{\bf{X}}_k}( {{{\bf{F}}_k},{{\{ {{q_{kj}}} \}}_{j \ne k}},{\mu _{kk}}}, {{\beta _k}}) \buildrel \Delta \over = \left[ {\begin{array}{*{20}{c}}
{{{\bf{F}}_k} + {\mu _{kk}}{\bf{I}}}&{{{\bf{F}}_k}{{\widehat {\bf{h}}}_{kk}}}\\
{\widehat {\bf{h}}_{kk}^H{{\bf{F}}_k}}&{\widehat {\bf{h}}_{kk}^H{{\bf{F}}_k}{{\widehat {\bf{h}}}_{kk}} + \sum\limits_{j = 1,j \ne k}^K {{q_{kj}}}  - \frac{{{\psi _k}}}{{{\xi _k}}}{\beta _k} + \sigma _k^2 - {\mu _{kk}}\eta _{kk}^2}
\end{array}} \right]\succeq {\bf{0}}
\end{equation}
\begin{equation} \label{EHcons2_SDP}
{{\bf{Y}}_{kj}}( {{{\bf{F}}_j},{q_{kj}},{\mu _{kj}}} ) \buildrel \Delta \over = \left[ {\begin{array}{*{20}{c}}
{{{\bf{F}}_j} + {\mu _{kj}}{\bf{I}}}&{{{\bf{F}}_j}{{\widehat {\bf{h}}}_{kj}}}\\
{\widehat {\bf{h}}_{kj}^H{{\bf{F}}_j}}&{\widehat {\bf{h}}_{kj}^H{{\bf{F}}_j}{{\widehat {\bf{h}}}_{kj}} - {q_{kj}} - {\mu _{kj}}\eta _{kj}^2}
\end{array}} \right] \succeq {\bf{0}},j \ne k
\end{equation}
\hrulefill \vspace{-1.2em}
\end{figure*}

\subsection{Proposed Rank-one Recovery Method} \label{sec_rankrecovery}
The matrices ${{\bf{F}}_k}$ obtained by solving the relaxed problem (\ref{robustSDP}) are not guaranteed to be of rank one in general, and hence, the solution provides a lower bound to the original problem (\ref{robustproblem}). If ${{\bf{F}}_k}$ happens to be of rank one, then the principal eigenvector ${\bf{f}}_k^ * $ of ${{\bf{F}}_k}$, such that ${\bf{F}}_k = {\bf{f}}_k^*{\bf{f}}_k^{*H}$ will be the optimal solution to problem (\ref{robustproblem}).\footnote{$\| {{\bf{f}}_k^*} \| = \sqrt {{f_k}}$, and $f_k$ is the largest eigenvalue of ${{\bf{F}}_k}$.} Otherwise, one has to resort to other techniques to obtain a suboptimal rank-one solution from ${{\bf{F}}_k}$. In this work, inspired by \cite{BeamformingforMISO}, we provide a simple heuristic, yet effective approach to overcome this difficulty when when higher-rank solutions are returned by solving problem (\ref{robustSDP}).\footnote{We can also use randomization techniques \cite{Semidefiniterelaxation} to recover a rank one solution.}

Before we proceed to introduce the rank-one recovery method, we first calculate the worst-case channels for given beamforming vectors and PS ratios.
Assuming that $\left\{ {{\bf{f}}_k^*} \right\}$ (the  principal eigenvectors of $\{{\bf{F}}_k\}$) and $\left\{ {{\rho _k^*}} \right\}$ have been determined in the previous subsection, then the worst-case CSI errors which minimize the SINR of user $k$, are the solutions to the following problems \\[-2mm]
\begin{equation}  \label{worstSINR1} \\[-2mm]
{\mathop {\min }\limits_{\{ {{{\bf{e}}_{kk}}} \}} \;| {(\widehat {\bf{h}}_{kk}^H + {\bf{e}}_{kk}^H){\bf{f}}_k^*} |\quad \begin{array}{*{20}{c}}
{{\textrm{s}}.{\textrm{t}}.}&{ \|{\bf{e}}_{kk}\|^2 \le \eta _{kk}^2,}
\end{array}}
\end{equation}
\begin{equation} \label{worstSINR2} \\[-2mm]
{\mathop {\max }\limits_{\{ {{{\bf{e}}_{kj}}} \}} \;| {(\widehat {\bf{h}}_{kj}^H + {\bf{e}}_{kj}^H){\bf{f}}_j^*} |\quad \begin{array}{*{20}{c}}
{{\textrm{s}}.{\textrm{t}}.}&{\|{\bf{e}}_{kj}\|^2 \le \eta _{kj}^2,\;j \ne k.}
\end{array}}
\end{equation}
\\[-2mm]
To solve the constrained optimization problem (\ref{worstSINR1}), we resort to the following inequality for any complex number $x$ and $y$ \\[-2mm] 
\begin{equation} \label{inq_1} \\[-2mm]
\left| {\left| x \right| - \left| y \right|} \right| \le \left| {x + y} \right| \le \left| {\left| x \right| + \left| y \right|} \right|.
\end{equation}
According to (\ref{inq_1}), we have the following inequality \\[-1mm]
\begin{equation} \label{inequality2} \\[-1mm]
\begin{array}{l}
| {{\bf{h}}_{kk}^H{{\bf{f}}_k}} | = | {\widehat {\bf{h}}_{kk}^H{{\bf{f}}_k} + {\bf{e}}_{kk}^H{{\bf{f}}_k}} |\\
 \ge | {| {\widehat {\bf{h}}_{kk}^H{{\bf{f}}_k}} | - | {{\bf{e}}_{kk}^H{{\bf{f}}_k}} |} | \ge | {| {\widehat {\bf{h}}_{kk}^H{{\bf{f}}_k}} | - {\eta _{kk}}\| {{{\bf{f}}_k}} \|} |,
\end{array}
\end{equation}
where the second inequality follows from the boundedness of CSI errors and the Cauchy-Schwartz inequality. It is worth noting that we have used a mild assumption in (\ref{inequality2}) that $| {\widehat {\bf{h}}_{kk}^H{{\bf{f}}_k}} | \geq | {{\bf{e}}_{kk}^H{{\bf{f}}_k}} |$, which holds true in most cases since the CSI error is often much smaller than the channel coefficients and ${\bf{h}}_{kk}$ cannot be orthogonal to ${{\bf{f}}_k}$.
We observe that the first inequality in (\ref{inequality2}) holds with equality if and only if ${\bf{e}}_{kk}^H{{\bf{f}}_k^*} = \alpha \widehat {\bf{h}}_{kk}^H{{\bf{f}}_k^*},\,\alpha  \in {\mathbb{R} _ - }$. Also the second inequality in (\ref{inequality2}) holds with equality if and only if ${\bf{e}}_{kk}^{} = \beta {{\bf{f}}_k^*},\,\beta  \in {\mathbb{C}}$. Thus the optimal solution $\overline {\bf{e}} _{kk}^{}$ of (\ref{worstSINR1}) can be obtained as \\[-2mm]
\begin{equation} \label{worst1} \\[-1mm]
\alpha  = \frac{{ - {\eta _{kk}}{{{\bf{f}}_k^{*H}}}{\bf{f}}_k^*}}{{\|{\bf{f}}_k^*\| |\widehat {\bf{h}}_{kk}^H{\bf{f}}_k^*|}},\;\beta  = {\textrm{conj}}\left( {\frac{{ - {\eta _{kk}}\widehat {\bf{h}}_{kk}^H{\bf{f}}_k^*}}{{\|{\bf{f}}_k^*\| |\widehat {\bf{h}}_{kk}^H{\bf{f}}_k^*|}}} \right),
\end{equation}
\begin{equation} \label{worst3} \\[-1mm]
\overline {\bf{e}} _{kk}^{} = \beta {\bf{f}}_k^*.
\end{equation}
Proceeding in the same manner, the optimal solution $\overline {\bf{e}} _{kj}^{}$ of (\ref{worstSINR2}) can be obtained as \\[-2mm]
\begin{equation} \\[-1mm]
\alpha  = \frac{{{\eta _{kj}}{\bf{f}}{{_j^*}^H}{\bf{f}}_j^*}}{{\|{\bf{f}}_j^*\| |\widehat {\bf{h}}_{kj}^H{\bf{f}}_j^*|}},\; \beta  = {\textrm{conj}}\left( {\frac{{{\eta _{kj}}\widehat {\bf{h}}_{kj}^H{\bf{f}}_j^*}}{{\|{\bf{f}}_j^*\| |\widehat {\bf{h}}_{kj}^H{\bf{f}}_j^*|}}} \right),
\end{equation}
\begin{equation} \\[-1mm]
\overline {\bf{e}} _{kj}^{} = \beta {\bf{f}}_j^*.
\end{equation}

Similarly, the worst-case CSI errors which minimize the EH of user $k$ are the solution to the following problem \\[-2mm]
\begin{equation} \label{worstEH2} \\[-1mm]
{\mathop {\min }\limits_{\{ {{\bf{e}}_{kj}}\} } \;| {(\widehat {\bf{h}}_{kj}^H + {\bf{e}}_{kj}^H){\bf{f}}_j^*} |\; \begin{array}{*{20}{c}}
{{\textrm{s}}.{\textrm{t}}.}\;{\|{\bf{e}}_{kj}\|^2}
\end{array} \le \eta _{kj}^2,\;\forall j,k \in \mathcal{K}.}
\end{equation}
Let $\widetilde {\bf{e}}_{kk}^{}$ denote the optimal solution of (\ref{worstEH2}) when $j=k$, which can be calculated following the same approach as used for (\ref{worst1}) and (\ref{worst3}). In the case ${j \ne k}$, we modify the cost function using the method of Lagrange multipliers \cite{AdaptiveFilterTheory}, which yields the following Lagrangian function\footnote{In this case, (\ref{worstEH2}) cannot be simply solved because the mild assumption does not hold in this case.} \\[-2mm]
\begin{equation} \label{Lagmulti} \\[-1mm]
{\mathcal{L}} = (\widehat {\bf{h}}_{kj}^H + {\bf{e}}_{kj}^H){\bf{F}}_j^*(\widehat {\bf{h}}_{kj}^{} + {\bf{e}}_{kj}^{}) + {\tau _{kj}}({\bf{e}}_{kj}^H{\bf{e}}_{kj}^{} - \eta _{kj}^2),
\end{equation}
where ${\bf{F}}_j^* = {\bf{f}}_j^*{( {{\bf{f}}_j^*} )^H}$ and ${\tau _{kj}}$ is the Lagrange multiplier associated with the bounded CSI error constraint from transmitter $j$ to receiver $k$. Taking the gradient of ${\mathcal L}$ in (\ref{Lagmulti}) with respect to $\textrm{conj}({{\bf{e}}_{kj}})$, we can obtain \\[-3mm]
\begin{equation} \label{wosrt-casechannel1} \\[-1mm]
\widetilde {\bf{e}}_{kj}^{} =  - {({\bf{F}}_j^* + {\tau _{kj}}{\bf{I}})^{ - 1}}{\bf{F}}_j^*\widehat {\bf{h}}_{kj}^{} = \frac{{{\bf{F}}_j^*\widehat {\bf{h}}_{kj}^{}}}{{{\tau _{kj}} + {{\| {{\bf{f}}_j^*} \|}^2}}}.
\end{equation}

The Lagrange multiplier ${\tau _{kj}}$ can be determined by solving \\[-3mm]
\begin{equation} \label{solveekk} \\[-1mm]
\widetilde {\bf{e}} _{kj}^H\widetilde {\bf{e}} _{kj}^{} - \eta _{kj}^2 = 0.
\end{equation}
Introducing ${{\bf{g}}_{kj}} = {\bf{F}}_j^*\widehat {\bf{h}}_{kj}^{}$, (\ref{solveekk}) becomes \\[-1mm]
\begin{equation} \\[-1mm]
\frac{{{\bf{g}}_{kj}^H{{\bf{g}}_{kj}}}}{{{{({\tau _{kj}} + {{\| {{\bf{f}}_j^*} \|}^2})}^2}}} = \eta _{kj}^2.
\end{equation}
Therefore, the lagrange multiplier ${{\tau _{kj}}}$ can be given by
\begin{equation} \label{Lagrangemultiplier1}
{\tau _{kj}} = \sqrt {{\bf{g}}_{kj}^H{{\bf{g}}_{kj}}/\eta _{kj}^2}  - {\| {{\bf{f}}_j^*} \|^2}.
\end{equation}



We note that $\widetilde {\bf{e}}_{kk}^{} = \overline {\bf{e}} _{kk}^{}$ but in the case $j \ne k$, $\widetilde {\bf{e}}_{kj}^{}$ and $\overline {\bf{e}} _{kj}^{}$ can not be simultaneously attained for the same channel realizations, which means that a CSI error vector minimizing both SINR and EH with given beamforming vectors and PS ratios does not exit in general. However, we employ both $\widetilde {\bf{e}}_{kj}^{}$ and $\overline {\bf{e}} _{kj}^{}$ to guarantee the robustness of the joint design.


With the worst-case analysis described above, we recover the rank-one solution by scaling up the beamforming vector $ {{\bf{f}}_k^*} $ by $ {\sqrt {{\varphi _k}} } $ and then jointly optimize $\{ {{\varphi _k}} \}$ and receive PS ratios $\{ {{\rho _k}} \}$ to satisfy both worst-case SINR and EH constraints and yet minimize the total transmission power. Specifically, we consider the following problem with given $\{ {{\bf{f}}_k^*} \}$
\begin{subequations} \label{recoverRANK1}
\begin{align}
\mathop {\min }\limits_{\{ {{\rho _k},\; {\varphi _k}} \}}  \;\; & \sum\limits_{k = 1}^K {{\varphi _k}{{\| {{\bf{f}}_k^*} \|}^2}} \\
\textrm{s.t.}\;\;\;\; & \frac{{{\rho _k}{\varphi _k}{u_{kk}}}}{{\sum\limits_{j = 1,j \ne k}^K {{\rho _k}{\varphi _j}{{\widetilde{u}_{kj}}}}  + {\rho _k}\sigma _k^2 + \omega _k^2}} \ge {\gamma _k},\label{recoverSINR1}\\
& {\xi _k}\left( {1 - {\rho _k}} \right)\left( {\sum\limits_{j = 1}^K {{\varphi _j}{u_{kj}}}  + \sigma _k^2} \right) \ge {\psi _k},\label{recoverEH1}\\
& 0 \le {\rho _k} \le 1,\;\;{\varphi _k} > 0,\;\;\forall k \in \mathcal{K},
\end{align}
\end{subequations}
where $\{{u_{kj}}\}_{j,k=1}^K = \{| {( {\widehat {\bf{h}}_{kj}^H + \widetilde {\bf{e}}_{kj}^H} ){\bf{f}}_j^*} |^2\}$ denote the minimum interference power and $\{\widetilde{u}_{kj}\}_{k=1,j\neq k}^K = \{| {( {\widehat {\bf{h}}_{kj}^H + \overline {\bf{e}} _{kj}^H} ){\bf{f}}_j^*} |^2\}$ denote the maximum interference power. Let us define ${x_k} = \frac{{{\varphi _k}}}{{{\gamma _k}}}{u_{kk}} - \sum\nolimits_{j = 1,j \ne k}^K {{\varphi _j}{{\widetilde u}_{kj}}}  - \sigma _k^2$, $x_k$ is implicitly supposed to be larger than $0$, and note that otherwise (\ref{recoverSINR1}) will be infeasible.
Thus, (\ref{recoverSINR1}) can be rewritten as \\[-2mm]
\begin{equation} \\[-2mm]
\| {{{\left[ {2{\omega _k},{x_k} - {\rho _k}} \right]}}} \| \le {x_k} + {\rho _k}.
\end{equation}
Similarly, by introducing ${y_k} = \sum\limits_{j = 1}^K {{\varphi _j}{u_{kj}}}  + \sigma _k^2$, (\ref{recoverEH1}) is equivalent to \\[-2mm]
\begin{equation}\\[-2mm]
\| {[ {2\sqrt {{\psi _k}/{\xi _k}} ,{y_k} + {\rho _k} - 1} ]} \| \le {y_k} - {\rho _k} + 1.
\end{equation}

Then problem (\ref{recoverRANK1}) can be reformulated as
\begin{equation} \label{recoverSOCP}
\begin{array}{l}
\mathop {\min }\limits_{\{ {\rho _k},\;{\varphi _k}\} } \;\;\sum\limits_{k = 1}^K {{\varphi _k}\|{\bf{f}}_k^*\|{^2}} \\
{\textrm{s}}.{\textrm{t}}.\;||[2{\omega _k},{x_k} - {\rho _k}]|| \le {x_k} - {\rho _k},\\
\| {[ {2\sqrt {{\psi _k}/{\xi _k}} ,{y_k} + {\rho _k} - 1} ]} \| \le {y_k} - {\rho _k} + 1,\\
{x_k} = \frac{{{\varphi _k}}}{{{\gamma _k}}}{u_{kk}} - \sum\limits_{j = 1,j \ne k}^K {{\varphi _j}{{\widetilde u}_{kj}}}  - \sigma _k^2,\\
{y_k} = \sum\limits_{j = 1}^K {{\varphi _j}{u_{kj}}}  + \sigma _k^2,\;0 \le {\rho _k} \le 1,\;\;\\
{\varphi _k} > 0,\;\;{x_k} \ge 0,\;{y_k} \ge 0,\;\forall k \in \mathcal{K}.
\end{array}
\end{equation}
\\[-2mm]
The above optimization problem is a SOCP problem \cite{Applicationsofsecond} because its objective function is linear and its constraints are linear or second-order cones. It can be efficiently solved by off-the-shelf algorithms.
The proposed SDR-based robust algorithm for problem (\ref{robustproblem}) is summarized in Table \ref{Algorithm1}.
\begin{table}
  \centering
  \caption{Algorithm-1 : Proposed SDR-based robust design}\label{Algorithm1}
  \begin{tabular}{p{0.9\columnwidth}}
 \hline
 \begin{itemize}
 \item[1.] Solve problem (\ref{robustSDP}) to obtain the optimal $\left\{ {{\bf{F}}_k^*} \right\}$ and $\left\{ {\rho _k^*} \right\}$.
 \item[2.] For each $k$, find the principal eigenvector $\left\{ {{\bf{f}}_k^*} \right\}$ of ${{\bf{F}}_k^*}$. If ${\textrm{rank}}\left( {{\bf{F}}_k^*} \right) = 1,\;\forall k \in \mathcal{K}$, then $\left\{ {{\bf{f}}_k^*} \right\}$ and $\left\{ {\rho _k^*} \right\}$ are the optimum solution and exit the algorithm, otherwise go to Step $3$.
 \item[3.] Solve problem (\ref{recoverSOCP}) with $\left\{ {{\bf{f}}_k^*} \right\}$ to obtain $\left\{ {{\varphi _k}} \right\}$ and $\left\{ {{\rho _k}} \right\}$.
 \item[4.] Return the beamforming vectors $\left\{ {\sqrt {{\varphi _k}} {\bf{f}}_k^*} \right\}$ and $\left\{ {{\rho _k}} \right\}$.
 \end{itemize} \\
  \hline
 \end{tabular}\vspace{-1.2em}
\end{table}

\subsection{SOCP Relaxation}  \label{SOCP_Relaxation}
In Subsection A, we relaxed the robust JBPS problem as a SDP problem. It is well known that solving a SDP problem requires relatively high computational complexity.
To obtain a low complexity solution, we propose to formulate the original problem as a SOCP problem based on proper relaxations, i.e. SOCP relaxation. Similar to \cite{JointBeamformingAnd}, problem (\ref{robustproblem}) can be relaxed as the following problem by replacing the EH constraints with the sum of the SINR and EH constraints \\[-2mm]
\begin{equation} \label{robust_relax_problem} \\[-1mm]
\begin{array}{*{20}{l}}
{\begin{array}{*{20}{c}}
{\mathop {\min }\limits_{\{ {{{\bf{f}}_k},{\rho _k}} \}} }&{\sum\limits_{k = 1}^K {{{\| {{{\bf{f}}_k}} \|}^2}} }
\end{array}}\\
{\begin{array}{*{20}{c}}
{{\textrm{s}}.{\textrm{t}}.}&{\begin{array}{*{20}{l}}
{\frac{1}{{{\gamma _k}}}{{| {{\bf{h}}_{kk}^H{{\bf{f}}_k}} |}^2} - \sum\limits_{j \ne k}^K {{{| {{\bf{h}}_{kj}^H{{\bf{f}}_j}} |}^2}}  \ge \sigma _k^2 + \frac{{\omega _k^2}}{{{\rho _k}}},}\\
{\left( 1 + \frac{1}{{{\gamma _k}}} \right){{| {{\bf{h}}_{kk}^H{{\bf{f}}_k}} |}^2} \ge \frac{{{\psi _k}}}{{{\xi _k}( {1 - {\rho _k}} )}} + \frac{{\omega _k^2}}{{{\rho _k}}},}\\
{0 \le {\rho _k} \le 1,\; \| {\bf{e}}_{kj}\|^2 \le \eta _{kj}^2,\;\forall j,k \in \mathcal{K}.}
\end{array}}
\end{array}}
\end{array}
\end{equation}

According to (\ref{inq_1}), we have the following inequality
\begin{equation} \label{inequality1} \\[-1mm]
\begin{array}{l}
| {{\bf{h}}_{kj}^H{{\bf{f}}_j}} | = | {\widehat {\bf{h}}_{kj}^H{{\bf{f}}_j} + {\bf{e}}_{kj}^H{{\bf{f}}_j}} |\\
 \le | {| {\widehat {\bf{h}}_{kj}^H{{\bf{f}}_j}} | + | {{\bf{e}}_{kj}^H{{\bf{f}}_j}} |} | \le | {| {\widehat {\bf{h}}_{kj}^H{{\bf{f}}_j}} | + {\eta _{kj}}\| {{{\bf{f}}_j}} \|} |,\;j\neq k.
\end{array}
\end{equation}
Together with inequality (\ref{inequality2}), the SINR constraints in (\ref{robust_relax_problem}) can be reformulated as \\[-2mm]
\begin{equation} \label{SINR_cons} \\[-1mm]
\begin{array}{l}
\frac{1}{{{\gamma _k}}}{| {| {\widehat {\bf{h}}_{kk}^H{{\bf{f}}_k}} | - {\eta _{kk}}\| {{{\bf{f}}_k}} \|} |^2}\\
 \ge \sum\limits_{j = 1,j \ne k}^K {{{| {| {\widehat {\bf{h}}_{kj}^H{{\bf{f}}_j}} | + {\eta _{kj}}\| {{{\bf{f}}_j}} \|} |}^2}}  + \sigma _k^2 + \frac{{\omega _k^2}}{{{\rho _k}}}.
\end{array}
\end{equation}
Similarly, the EH constraints in (\ref{robust_relax_problem}) can be expressed as \\[-1mm]
\begin{equation} \label{EH_cons}\\[-1mm]
\left( {1 + \frac{1}{{{\gamma _k}}}} \right){| {| {\widehat {\bf{h}}_{kk}^H{{\bf{f}}_k}} | - {\eta _{kk}}\| {{{\bf{f}}_k}} \|} |^2} \ge \frac{{{\psi _k}}}{{{\xi _k}( {1 - {\rho _k}} )}} + \frac{{\omega _k^2}}{{{\rho _k}}}.
\end{equation}

Furthermore, by introducing auxiliary variables ${\beta _{kj}}^\prime \textmd{s}$, which satisfy \\[-2mm]
\begin{equation}\\[-1mm]
| {\widehat {\bf{h}}_{kk}^H{{\bf{f}}_k}} | - {\eta _{kk}}\left\| {{{\bf{f}}_k}} \right\| \ge {\beta _{kk}},
\end{equation}
\begin{equation} \\[-1mm]
| {\widehat {\bf{h}}_{kj}^H{{\bf{f}}_j}} | + {\eta _{kj}}\left\| {{{\bf{f}}_j}} \right\| \le {\beta _{kj}}, \; j\neq k,
\end{equation}
(\ref{SINR_cons}) and (\ref{EH_cons}) can be expressed as\\[-2mm]
\begin{equation} \\[-1mm]
\frac{1}{{{\gamma _k}}}\beta _{kk}^2 \ge \sum\limits_{j = 1,j \ne k}^K {\beta _{kj}^2}  + \sigma _k^2 + \frac{{\omega _k^2}}{{{\rho _k}}},
\end{equation}
\begin{equation}
\left( {1 + \frac{1}{{{\gamma _k}}}} \right)\beta _{kk}^2 \ge \frac{{{\psi _k}}}{{{\xi _k}( {1 - {\rho _k}} )}} + \frac{{\omega _k^2}}{{{\rho _k}}}.
\end{equation}
Let $a_k^2 = {\rho _k}$ and $b_k^2 = 1 - {\rho _k}$, it follows that $a_k^2 + b_k^2 = 1$. By further introducing $c_k^2 \ge \frac{{\omega _k^2}}{{a_k^2}}$, $d_k^2 \ge \frac{{{\psi _k}}}{{{\xi _k}b_k^2}}$ and $| {\widehat {\bf{h}}_{kj}^H{{\bf{f}}_j}} | \le {e_{kj}}$, we can write (\ref{robust_relax_problem}) as the following SOCP problem
\begin{equation} \label{robust_SOCP}
\begin{array}{*{20}{l}}
{\mathop {\min }\limits_{\{ {{\bf{f}}_k},\;{a_k},\;{b_k},\;{c_k},\;{d_k},\;{\beta _{kj}}\},\;{\{ {e_{kj}}\} _{j \ne k}}} \quad t}\\
{{\textrm{s}}.{\textrm{t}}.\;\| {{{[ {{{ {{{\bf{f}}_1^T}} }}, \ldots ,{{ {{{\bf{f}}_K^T}} }}} ]}}} \| \le t,}\\
{\;\;\quad \| {{{[ {{\bf{J}}_k^T,{\sigma _k},{c_k}} ]}}} \| \le \frac{1}{{\sqrt {{\gamma _k}} }}{\beta _{kk}},}\\
{\;\;\quad {\eta _{kj}}\| {{{\bf{f}}_j}} \| \le  - e_{kj} + {\beta _{kj}},\;| {\widehat {\bf{h}}_{kj}^H{{\bf{f}}_j}} | \le e_{kj},\; j\neq k,}\\
{\;\;\quad {\eta _{kk}}\| {{{\bf{f}}_k}} \| \le \widehat {\bf{h}}_{kk}^H{{\bf{f}}_k} - {\beta _{kk}},}\\
{\;\;\quad \| {{{[ {{c_k},{d_k}} ]}}} \| \le \sqrt {1 + \frac{1}{{{\gamma _k}}}} {\beta _{kk}},\forall i,}\\
{\;\;\quad \|[2{({\psi _k}/{\xi _k})^{1/4}},{d_k} - {b_k}]\| \le {d_k} + {b_k},}\\
{\;\;\quad \| {{{[ {2\sqrt {{\omega _k}} ,{c_k} - {a_k}} ]}}} \| \le {c_k} + {a_k},}\\
{\;\;\quad \| {{{[ {{a_k},{b_k}} ]}}} \| \le 1,}\\
{\;\;\quad {a_k} \ge 0,\;{b_k} \ge 0,\;{\beta _{kj}} \ge 0,{\mkern 1mu} \forall j,k, \in \mathcal{K}}
\end{array}
\end{equation}
where \\[-2mm]
\begin{equation} \\[-1mm]
{\bf{J}}_k^{} = {[ {{\beta _{k1}}, \ldots {\beta _{k(k - 1)}},{\beta _{k(k + 1)}}, \ldots ,{\beta _{kK}}} ]^T}.
\end{equation}
The constraints ${\| {{{[ {{a_k},{b_k}} ]}}} \| \le 1}$ must be satisfied with equality at optimality; otherwise the objective value can be further decreased by increasing ${a_k}^\prime \textmd{s}$. Note that we can restrict $\widehat {\bf{h}}_{kk}^H{{\bf{f}}_k}$ to be positive, which incurs no loss of optimality since we can always phase-rotate the vector ${{\bf{f}}_k}$ such that $\widehat {\bf{h}}_{kk}^H{{\bf{f}}_k}$ is positive real without affecting the cost function or the constraints. Since the solution to the SOCP relaxation problem (\ref{robust_relax_problem}) may not be a feasible solution to problem (\ref{robustproblem}), a robust solution recovery method must be employed to ensure the robustness, which will be introduced in the following subsection.

\subsection{Proposed Closed-form Robust Solution Recovery Method}
\label{close-form_recovery}
In Subsection B, we introduced a rank-one recovery method to recover a robust solution for problem (\ref{robustSDP}) by solving a SOCP problem (\ref{recoverSOCP}). Due to the SOCP relaxation, the solution to problem (\ref{robust_SOCP}) may not be robust to all possible channel realizations. In this subsection, we propose a closed-form recovery method in which $\{ {{\bf{f}}_k^ * } \}$ obtained by solving (\ref{robust_SOCP}) are scaled up by a common factor ${\sqrt \varphi  }$, and then jointly optimize $\varphi $ and the receive PS ratios $\{ {{\rho _k}} \}$ to satisfy both the worst-case SINR and EH constraints. Since we relax the EH constraints in (\ref{robust_SOCP}), it is required that $\varphi >1$ to ensure the robustness of the algorithm, which means that more power is needed to satisfy the worst-case EH constraints. It is worth noting that this closed-form method was first proposed in \cite{JointTransmitBeamforming} where perfect CSI is considered.

Similar to problem (\ref{recoverRANK1}), we consider the following problem with given $\left\{ {{\bf{f}}_k^*} \right\}$
\\[-2mm]
\begin{equation} \label{recoverRANK2}
\begin{array}{l}
\mathop {\min }\limits_{\varphi ,\;\{ {\rho _k}\} } \quad \varphi \sum\limits_{k = 1}^K {\|{\bf{f}}_k^*\|{^2}} \\
{\textrm{s}}.{\textrm{t}}.\;\frac{{\varphi {\rho _k}{u_{kk}}}}{{\varphi \sum\limits_{j = 1,j \ne k}^K {{\rho _k}{{\widetilde u}_{kj}}}  + {\rho _k}\sigma _k^2 + \omega _k^2}} \ge {\gamma _k},\\
{\xi _k}(1 - {\rho _k})\left( {\varphi \sum\limits_{j = 1}^K {{u_{kj}}}  + \sigma _k^2} \right) \ge {\psi _k},\\
0 \le {\rho _k} \le 1,\;\;\varphi  > 1,\;\;\forall k \in \mathcal{K},
\end{array}
\end{equation}
\\[-2mm]
where the definitions of ${{u_{kj}}}$ and ${{{\widetilde{u}_{kj}}}}$ have already been stated in Section \ref{sec_rankrecovery}. Introducing ${\widetilde x_k} = \frac{1}{{{\gamma _k}}}{u_{kk}} - \sum\limits_{j = 1,j \ne k}^K {{{\widetilde{u}_{kj}}}} $ and ${\widetilde y_k} = \sum\limits_{j = 1}^K {{u_{kj}}} ,\;\forall k \in \mathcal{K}$, problem (\ref{recoverRANK2}) can be equivalently rewritten as
\\[-2mm]
\begin{equation} \label{simplerecovery}
\begin{array}{l}
\mathop {\min }\limits_{\varphi,\;\{ {{\rho _k}} \}} \quad \varphi \\
\textrm{s.t.}\; {\rho _k} \ge \frac{{\omega _k^2}}{{\varphi {{\widetilde x}_k} - \sigma _k^2}},\\
1 - {\rho _k} \ge \frac{{{\psi _k}}}{{{\xi _k}\left( {\varphi {{\widetilde y}_k} + \sigma _k^2} \right)}},\\
0 \le {\rho _k} \le 1,\;\varphi  \ge 1,\;\forall k \in \mathcal{K}.
\end{array}
\end{equation}
Similar to \cite{JointTransmitBeamforming}, problem (\ref{simplerecovery}) admits a closed-form solution, which is given by
\begin{equation} \label{simplerecovery2} \\[-2mm]
\begin{array}{*{20}{l}}
{\mathop {\min }\limits_{\varphi  \ge 1} \quad \varphi }\\
{{\textrm{s}}.{\textrm{t}}.\quad {{\widetilde \varphi }_k} = \left\{ {\begin{array}{*{20}{l}}
&{1,\quad 0 < {g_k}( 1 ) \le 1}\\
&{\bar \varphi ,\quad {g_k}( 1 ) > 1\;\textrm{or}\;{g_k}( 1 ) \le 0}
\end{array}} \right.,\;}\\
{\quad \quad \;\varphi  \ge {{\widetilde \varphi }_k},\;\forall k \in \mathcal{K},}
\end{array}
\end{equation}
where ${g_k}( \varphi  ) = \frac{{\omega _k^2}}{{\varphi {{\widetilde x}_k} - \sigma _k^2}} + \frac{{{\psi _k}}}{{{\xi _k}( {\varphi {{\widetilde y}_k} + \sigma _k^2} )}}$ and ${\overline \varphi  }$ is the largest root of the equation ${g_k}( \varphi  ) = 1$. Additionally, $\rho _k = \frac{{\omega _k^2}}{{{\varphi ^*}{{\widetilde x}_k} - \sigma _k^2}}$ is the corresponding PS ratio with given ${{\varphi ^*}}$, where ${{\varphi ^*}}$ is the optimal solution of (\ref{simplerecovery2}).
The proposed robust design with SOCP relaxation for problem (\ref{robustproblem}) is summarized in Table \ref{Algorithm2}.

\emph{Remark:} It is worth noting that the two algorithms introduced in Section \ref{sec_rankrecovery} and \ref{close-form_recovery} can both be employed to recover a feasible rank-one solution to the relaxed problem (\ref{robustSDP}) and (\ref{robust_SOCP}), respectively. The two algorithms exhibit a tradeoff between recovery accuracy and computational complexity. In \emph{Algorithm-1}, we employ the rank-one recovery method from Section \ref{sec_rankrecovery} while in \emph{Algorithm-2}, the recovery method from Section \ref{close-form_recovery} is employed to further reduce the computational complexity. It is also important to note that the two algorithms have very close performance in our simulations.
\begin{table}
  \centering
  \caption{Algorithm-2 : Proposed Robust Design with SOCP Relaxation} \label{Algorithm2}
  \begin{tabular}{p{0.9\columnwidth}}
 \hline
 \begin{itemize}
 \item[1.] Solve problem (\ref{robust_SOCP}) to obtain the optimal $\{ {{\bf{f}}_k^*} \}$ and $\{ {\rho _k^*} \}$.
 \item[2.] Solve problem (\ref{simplerecovery2}) with $\{ {{\bf{f}}_k^*} \}$ to obtain ${{\varphi ^*}}$ and $\{ {{\rho _k}} \}$.
 \item[3.] Return the beamforming vectors $\{ {\sqrt {{\varphi}^*} {\bf{f}}_k^*} \}$ and $\{ {{\rho _k}} \}$.
 \end{itemize} \\
  \hline
 \end{tabular}\vspace{-1.8em}
\end{table}
\section{Proposed Iterative Robust Design} \label{Section_iterative}
Since the robust design approaches presented in Section \ref{Section_SDP} rely on relaxation techniques, they may not guarantee the optimal solution, we propose an iterative algorithm based on CCCP \cite{yuille2003concave, lanckriet2009convergence} in this section to improve the relaxation solutions and possibly obtain the optimal solutions.

\subsection{CCCP-based Iterative Robust Design}
The CCCP technique is widely adopted for solving non-convex
problems \cite{lanckriet2009convergence} by transforming them into a sequence of convex programming problems. The basic idea behind the proposed CCCP-based algorithm is to iteratively approximate the original non-convex feasible set in (\ref{SDRproblem}) around the current solution by a convex subset and then solve the resulting convex approximation at each iteration \cite{cheng2012joint}.

While the conventional CCCP considers scalar functions, for the current application we need to extend it to the case of matrix functions. Our iterative method is motivated by the observation that the non-convex constraint $\textrm{rank}({{\bf{F}}_k}) = 1$ is equivalent to the constraint
\begin{equation} \label{rankone_constraint}
{{\bf{f}}_k}{\bf{f}}_k^H = {{\bf{F}}_k}.
\end{equation}
Hence, relaxing constraint (\ref{rankone_constraint}) to ${{\bf{f}}_k}{\bf{f}}_k^H \succeq {{\bf{F}}_k}$ results in the following problem
\begin{subequations} \label{iterative_SDP3}
\begin{align}
&{\mathop {\min }\limits_{\{ {{\bf{F}}_k},\;{{\bf{f}}_k},\;{\alpha _k},\;{\beta _k},\;{\lambda _{kj}},\;{\mu _{kj}}\},\;\{{p_{kj}},\;{q_{kj}}\}_{j \neq k},\;t } } \;{t^2}\\
&{\textrm{s}}.{\textrm{t}}.\;\; {\| {{{[ {{{{{{\bf{f}}_1^T}}}}, \ldots ,{{{{{\bf{f}}_K^T}}}}} ]}}} \| \le t,}\\
&(\ref{SINRcons1_SDP}), (\ref{SINRcons2_SDP}), (\ref{EHcons1_SDP})\; \text{and} \; (\ref{EHcons2_SDP}),\\
&{{\alpha _k} \ge 1,\;{\beta _k} \ge 1,}\;{{\textrm{invp}}( {{\alpha _k}} ) + {\textrm{invp}}( {{\beta _k}} ) \le 1,}\\
&p_{kj} \geq 0,\; q_{kj} \geq 0,\; j \neq k,\\
&{{\bf{f}}_k}{\bf{f}}_k^H\succeq{{\bf{F}}_k},\label{rankone_constraint_relax} \\
&{{{\bf{F}}_k}\succeq{\bf{0}},\;{\lambda _{kj}} \ge 0,\;{\mu _{kj}} \ge 0,\;\forall j,k \in \mathcal{K}.}
\end{align}
\end{subequations}
\newtheorem{lemma}{\underline{Lemma}}
\begin{lemma} \label{lamma1}
Problem (\ref{SDRproblem}) is equivalent to problem (\ref{iterative_SDP3}).
\end{lemma}

\emph{Proof:} As we can see, problem (\ref{iterative_SDP3}) is a relaxed version of problem (\ref{SDRproblem}), thus if the optimal solution $\{{\bf{f}}_k^*,{\bf{F}}_k^*,\rho_k^*\}$ of (\ref{iterative_SDP3}) satisfies ${\bf{f}}_k^*{\bf{f}}_k^{*H} = {\bf{F}}_k^*$, we can assert that problem (\ref{SDRproblem}) is equivalent to problem (\ref{iterative_SDP3}). We prove ${\bf{f}}_k^*{\bf{f}}_k^{*H} = {\bf{F}}_k^*$ by showing that $\textrm{rank}({\bf{F}}_k^*) = 1$, which is quite obvious since ${\bf{f}}_k^*{\bf{f}}_k^{*H}$ is a rank one matrix. This completes the proof.

As we can see, only constraint (\ref{rankone_constraint_relax}) in problem (\ref{iterative_SDP3}) is non-convex. Thus, with the help of the following inequality
\begin{equation} \label{CCCP_5}
\begin{array}{l}
{{\bf{f}}_k}{\bf{f}}_k^H - ( {{\bf{f}}_k^i{{( {{{\bf{f}}_k} - {\bf{f}}_k^i} )}^H} + ( {{{\bf{f}}_k} - {\bf{f}}_k^i} ){\bf{f}}_k^{iH} + {\bf{f}}_k^i{\bf{f}}_k^{iH}} )\\
 = ( {{{\bf{f}}_k} - {\bf{f}}_k^i} ){( {{{\bf{f}}_k} - {\bf{f}}_k^i} )^H}\succeq{\bf{0}},
\end{array}
\end{equation}
we can transform (\ref{rankone_constraint_relax}) to
\begin{equation} \label{CCCP_4}
{{\bf{f}}_k}{\bf{f}}_k^H\succeq{\bf{f}}_k^i{( {{{\bf{f}}_k} - {\bf{f}}_k^i} )^H} + ( {{{\bf{f}}_k} - {\bf{f}}_k^i} ){\bf{f}}_k^{iH} + {\bf{f}}_k^i{\bf{f}}_k^{iH}\succeq{{\bf{F}}_k},
\end{equation}
where ${\bf{f}}_k^i$ denotes the current feasible point in the $i$th iteration. 

With (\ref{CCCP_4}), problem (\ref{iterative_SDP3}) can be reformulated as the following convex optimization problem in the $i$th iteration of the proposed CCCP-based algorithm
\begin{subequations} \label{iterative_SDP4}
\begin{align}
&{\mathop {\min }\limits_{\{ {{\bf{F}}_k},\;{{\bf{f}}_k},\;{\alpha _k},\;{\beta _k},\;{\lambda _{kj}},\;{\mu _{kj}}\},\;\{{p_{kj}},\;{q_{kj}}\}_{j \neq k},\;t } }\;{P({\bf{f}}_k^i)}\\
&{\textrm{s}}.{\textrm{t}}.\;\; {\| {{{[ {{{{{{\bf{f}}_1^T}} }}, \ldots ,{{ {{{\bf{f}}_K^T}} }}} ]}}} \| \le t,} \label{power_constraint}\\
&(\ref{SINRcons1_SDP}), (\ref{SINRcons2_SDP}), (\ref{EHcons1_SDP})\; \text{and} \; (\ref{EHcons2_SDP}),\\
&{{\alpha _k} \ge 1,\;{\beta _k} \ge 1,}\\
&{{\textrm{invp}}( {{\alpha _k}} ) + {\textrm{invp}}( {{\beta _k}} ) \le 1,}\label{lamma2_2}\\
&p_{kj} \geq 0,\; q_{kj} \geq 0,\; j \neq k,\\
&{{\bf{f}}_k^i{{( {{{\bf{f}}_k} - {\bf{f}}_k^i} )}^H} + ( {{{\bf{f}}_k} - {\bf{f}}_k^i}){\bf{f}}_k^{iH} + {\bf{f}}_k^i{\bf{f}}_k^{iH} \succeq {{\bf{F}}_k},} \label{lamma2_3}\\
&{{{\bf{F}}_k}\succeq{\bf{0}},\;{\lambda _{kj}} \ge 0,\;{\mu _{kj}} \ge 0,\;\forall j,k \in \mathcal{K},}
\end{align}
\end{subequations}
where $P({\bf{f}}_k^i) = {t^2}$.

\begin{lemma} \label{lamma2}
Suppose problem (\ref{iterative_SDP4}) is feasible, then strong duality holds true for problem (\ref{iterative_SDP4}) and its dual problem.
\end{lemma}
The proof is relegated to Appendix \ref{appendixB}.

To summarize, we can see that the feasible set of problem (\ref{iterative_SDP3}) is a subset of the original set defined in problem (\ref{robustSDP}). Then, if the initial point $\{{\bf{f}}_k^0\}$ is feasible for (\ref{iterative_SDP3}), all the feasible points $\{{\bf{f}}_k^i\}$ obtained by iteratively solving problem (\ref{iterative_SDP3}) always belong to the true feasible set of (\ref{robustSDP}). The proposed CCCP-based iterative robust design for problem (\ref{robustproblem}) is summarized in Table \ref{Algorithm3}. Regarding its convergence, we have the following Lemma.

\begin{lemma} \label{lamma3}
\emph{Algorithm-3} produces a non-increasing sequence of objective values. Moreover, every limit point $\{{\bf{f}}_k^*\}$ of the iterates generated by \emph{Algorithm 3} is a KKT point of problem (\ref{SDRproblem}).
\end{lemma}

\emph{Proof} : Please see Appendix \ref{appendixC}.

\begin{table}
  \centering
  \caption{Algorithm-3 : Proposed CCCP-based Iterative Robust Design} \label{Algorithm3}
  \begin{tabular}{p{0.9\columnwidth}}
 \hline
 \begin{itemize}
 \item[1.] Define the tolerance of accuracy $\delta $ and the maximum iteration number ${N^{\max}}$. $\mathbf{Initialize}$ the algorithm with a feasible point $\{ {\bf{f}}_k^0,{\rho _k}\}$. Set the iteration number $i=0$.
 \item[2.] $\mathbf{Repeat}$:
 \begin{itemize}
 \item  Solve problem (\ref{iterative_SDP4}) with the current feasible point $\{ {\bf{f}}_k^i,{\rho _k}\} $.
 \item  Assign the solution to $\{ {\bf{f}}_k^{i+1},{\rho _k}\} $ and update the iteration number : $i = i + 1$.
 \end{itemize}
 \item[3.] $\mathbf{Until}$: the objective function converges, i.e. $|P^{i+1}-P^{i}| < \delta$ or the maximum number of iterations is reached, i.e. $i>{N^{\max}}$.
 \end{itemize} \\
  \hline
 \end{tabular} \vspace{-1.8em}
\end{table}
\subsection{The Proposed Initialization Method}
The proposed \emph{Algorithm-3} presented in the previous subsection requires an initial feasible point of problem (\ref{robustproblem}) \cite{yuille2003concave}. If problem (\ref{iterative_SDP4}) is initialized with an infeasible point, then the CCCP may fail at the first iteration.

The proposed initialization method is based on \emph{Algorithm-2}. As has been stated in Section \ref{SOCP_Relaxation}, the solution obtained by solving (\ref{robust_SOCP}) may not be robust to all channel realizations due to SOCP relaxation. In order to make the initialization method as simple as possible, we propose a new method instead of the robust solution recovery method in Section \ref{close-form_recovery}. We first observe that the solution obtained by solving (\ref{robust_SOCP}) can provide guaranteed SINR levels if (\ref{robust_SOCP}) is feasible. Thus, if we scale up the beamforming vector $\{ {\bf{f}}_k^*\} $ by a sufficiently large common factor $\sqrt \varphi  $, then $\{ \sqrt \varphi  {\bf{f}}_k^*,{\rho _k^*}\} $ will be a feasible point of problem (\ref{robustproblem}), since
\begin{equation}
\begin{array}{l}
\frac{{\varphi \rho _k^*{{| {(\widehat {\bf{h}}_{kk}^H + {\bf{e}}_{kk}^H){\bf{f}}_k^*} |}^2}}}{{\varphi \rho _k^*\left( {\sum\limits_{j = 1,j \ne k}^K {{{| {(\widehat {\bf{h}}_{kj}^H + {\bf{e}}_{kj}^H){\bf{f}}_j^*} |}^2}}  + \sigma _k^2} \right) + \omega _k^2}}\\
 \ge \frac{{\rho _k^*{{| {(\widehat {\bf{h}}_{kk}^H + {\bf{e}}_{kk}^H){\bf{f}}_k^*} |}^2}}}{{\rho _k^*\left( {\sum\limits_{j = 1,j \ne k}^K {{{| {(\widehat {\bf{h}}_{kj}^H + {\bf{e}}_{kj}^H){\bf{f}}_j^*} |}^2}}  + \sigma _k^2} \right) + \omega _k^2}},\;\forall \varphi  \ge 1.
\end{array}
\end{equation}
Let
\\[-2mm]
\begin{equation}
P_k^{*\textrm{EH}} = {\xi _k}( {1 - \rho _k^*} )\left( {\sum\limits_{j = 1}^K {{{| {(\widehat {\bf{h}}_{kj}^H + \widetilde {\bf{e}}_{kj}^H){\bf{f}}_j^*} |}^2}}  + \sigma _k^2} \right),
\end{equation}
\\[-2mm]
where $\widetilde {\bf{e}}_{kj}^H$ is calculated by (\ref{wosrt-casechannel1}) and (\ref{Lagrangemultiplier1}). Then $\varphi$ can be expressed as
\begin{equation}
\varphi  = \frac{{\frac{{{\psi _k}}}{{{\xi _k}( {1 - \rho _k^*} )}} - \sigma _k^2}}{{\frac{{P_k^{*\textrm{EH}}}}{{{\xi _k}( {1 - \rho _k^*} )}} - \sigma _k^2}}.
\end{equation}

We remark that \emph{Algorithm-3} consists of a two-stage algorithm for solving problem (\ref{robustproblem}). In the first stage, the initialization method is applied to find a feasible solution of problem (\ref{robustproblem}). If the initialization method fails to find a feasible solution, \emph{Algorithm-3} declares failure and stops. In the second stage, we iteratively solve problem (\ref{iterative_SDP4}).

\section{Complexity Analysis} \label{section:complexity}
In this section, we compare the relative computational complexities of the proposed robust design algorithms. As will be seen from our analysis and the simulation results in Section \ref{section:simulation}, the proposed robust algorithms exhibit a tradeoff between computational efficiency and robust performance. For ease of comparison, we assume that all transmitters are equipped with the same number of antennas, i.e., ${N_k} = N,\;\forall k \in \mathcal{K}$. Moreover, we apply the basic elements of complexity analysis as used in \cite{Wang2014Outage}.

\emph{1) Algorithm-1:}
Consider problem (\ref{robustSDP}), which involves $2K^2$ linear matrix inequality (LMI) constraints of size $N + 1$ and $K$ LMI constraints of size $N$.\footnote{Here, we ignore the constraints of lower sizes, since they will not affect the order of the whole problem.} Here, the number of decision variables $n$ is on the order of
$\mathcal{O} (KN^2 + 4K^2 )$. Thus, the complexity of a generic interior-point method (IPM) for solving problem (\ref{robustSDP}) is on the order of the quantity shown on the first row of Table \ref{complexity}.
Regarding the complexity of the rank-one recovery method (\ref{recoverSOCP}), we note that it involves $2K$ variables and $2K$ second-order cone (SOC) constraints of dimension $3$. Hence, it follows that the complexity of the SOCP problem is
$\mathcal{O} ( {2K \sqrt {4K}\,[ {2K {3^2} + 4{K^2}} ]} )$.
We note that the complexity of the worst-case channel vector calculation is dominated by singular value decomposition and matrix inversion operations, and it is considered to be negligible compared to solving problem (\ref{robustSDP}).

\emph{2) Algorithm-2:}
Problem (\ref{robust_SOCP}) involves $2K^2+4K+1$ SOC constraints, including $1$ SOCs of dimension $KN+1$, $K$ SOCs of dimension $K+2$, $K^2$ SOCs of dimension $N+1$, $4K$ SOCs of dimension $3$, and $K^2-K$ SOCs of dimension $2$. The number of variables $n$ is on the order of $\mathcal{O} ( KN + 3K + 2K^2 )$. Thus, the complexity of a generic IPM for solving problem (\ref{robust_SOCP}) is the quantity shown on the second row of Table \ref{complexity}.
Here, the complexity of the closed-form robust solution recovery method is dominated by the worst-case channel vector calculation and solving a quadratic equation, which is negligible.

\emph{3) Algorithm-3:} Problem (\ref{iterative_SDP4}) involves $KN^2 + KN + 4K^2+2K + 1$ variables, $1$ SOC constraints of size $KN+1$, $2K^2$ LMI constraints of size $N+1$, and $2K$ LMI constraints of size $N$. Hence, the complexity of the CCCP-based iterative robust design is as shown on the third row of Table \ref{complexity}.

We can check the asymptotic complexity of the proposed algorithms when $N$ and $K$ are large, i.e. let $N = K \rightarrow \infty $. One can verify that the complexities of the proposed algorithms in Table \ref{complexity} are on the orders of $2\sqrt{2}N^{11.5}$, $66N^7$ and $2\sqrt{2}N^{max}N^{11.5}$, respectively. As seen, the lowest complexity is achieved by \emph{Algorithm-2}, followed by \emph{Algorithm-1} and \emph{Algorithm-3}.
%
\begin{table*}
  \centering
  \caption{Complexity analysis of the robust designs} \label{complexity}
  \begin{tabular}{|c|c|c|c|c|c|c}

\hline \hline
   \multicolumn{1}{|c}{Robust design} & \multicolumn{1}{|c|}{Complexity Order (suppressing the $\ln ( {{1 \mathord{\left/
 {\vphantom {1 \varepsilon }} \right.
 \kern-\nulldelimiterspace} \varepsilon }} )$ )}\\ \hline \hline
   \multirow{2}*{Algorithm-1} & {$\mathcal{O} ( {n\sqrt {2{K^2}( {N + 1} ) + KN}\, [ {2{K^2}{{( {N + 1} )}^3} + K{N^3} + 2n{K^2}{{( {N + 1} )}^2} + nK{N^2} + {n^2}} ]} )$} \\
   & $ + {\mathcal{O}}( {2K\sqrt {4K}\,[ {2K {3^2} + 4{K^2}} ]} ),\;n = {\mathcal{O}}( {K{N^2} + 4{K^2}} )$ \\ \hline

   \multirow{2}*{Algorithm-2} & {$\mathcal{O} ( {n\sqrt {4K^2+8K+2}\,[ {(KN + 1)^2} + K{(K + 2)^2} + {K^2}{(N + 1)^2} + 4K {3^2} + ({K^2} - K) {2^2} + {n^2} ]} )$} \\
   & $n = {\mathcal{O}}( KN + 3K + 2K^2 )$ \\ \hline

   \multirow{2}* {Algorithm-3} & {The complexity of \emph{Algorithm-2} + ${\mathcal{O}}(nN^{max}\sqrt {2{K^2}(N + 1) + 2KN + 2}\,[2{K^2}{( {N + 1} )^3} + 2K{N^3} $ }\\
   & $ + 2n{K^2}{( {N + 1} )^2} + nK{N^2} + {(KN + 1)^2} + {n^2}])$, $ n=\mathcal{O} (KN^2 + KN+4K^2+2K+1) $, $N^{max}$ is the iteration number\\ \hline
 \end{tabular} \vspace{-1.8em}
\end{table*}

\section{Simulation Results} \label{section:simulation}
In this section, we evaluate the performance of the proposed robust JBPS algorithms numerically. We assume there are $K=3$ transmit-receive pairs and all transmitters are equipped with ${N_k} = N,\;k \in \{1,2,3\}$ antennas unless otherwise specified. We assume that each transmit-receive pair has the same set of parameters, i.e., ${\gamma _k} = \gamma $, ${\psi _k} = \psi $, ${\xi _k} = \xi $, $\sigma _k^2 = {\sigma ^2}$, $\omega _k^2 = {\omega ^2}$ and ${\eta _{kj}} = \eta ,\;\forall j,k \in \mathcal{K}$ for simplicity. Moreover, the nominal channel vectors $\{ {{{\widehat {\bf{h}}}_{kj}}} \}$ are randomly generated from independent and identical Rayleigh fading distribution with average power $1$. We set $\xi = 1$, ${\sigma ^2} = -30$ dBm, ${\omega ^2} = -20$ dBm, $\delta =10^{-4}$ and ${N^{\max}}=20$ in all our simulations. All the modelling and solution of the algorithms are performed using CVX \cite{CVX} on a desktop Intel (i3-2100) CPU running at 3.1GHz and 4GB RAM.

\begin{figure}[hbtp]
  \setlength{\abovecaptionskip}{-0.2cm} 
  \setlength{\belowcaptionskip}{-0.2cm} 
  \centering
  \includegraphics[width = 0.42\textwidth]{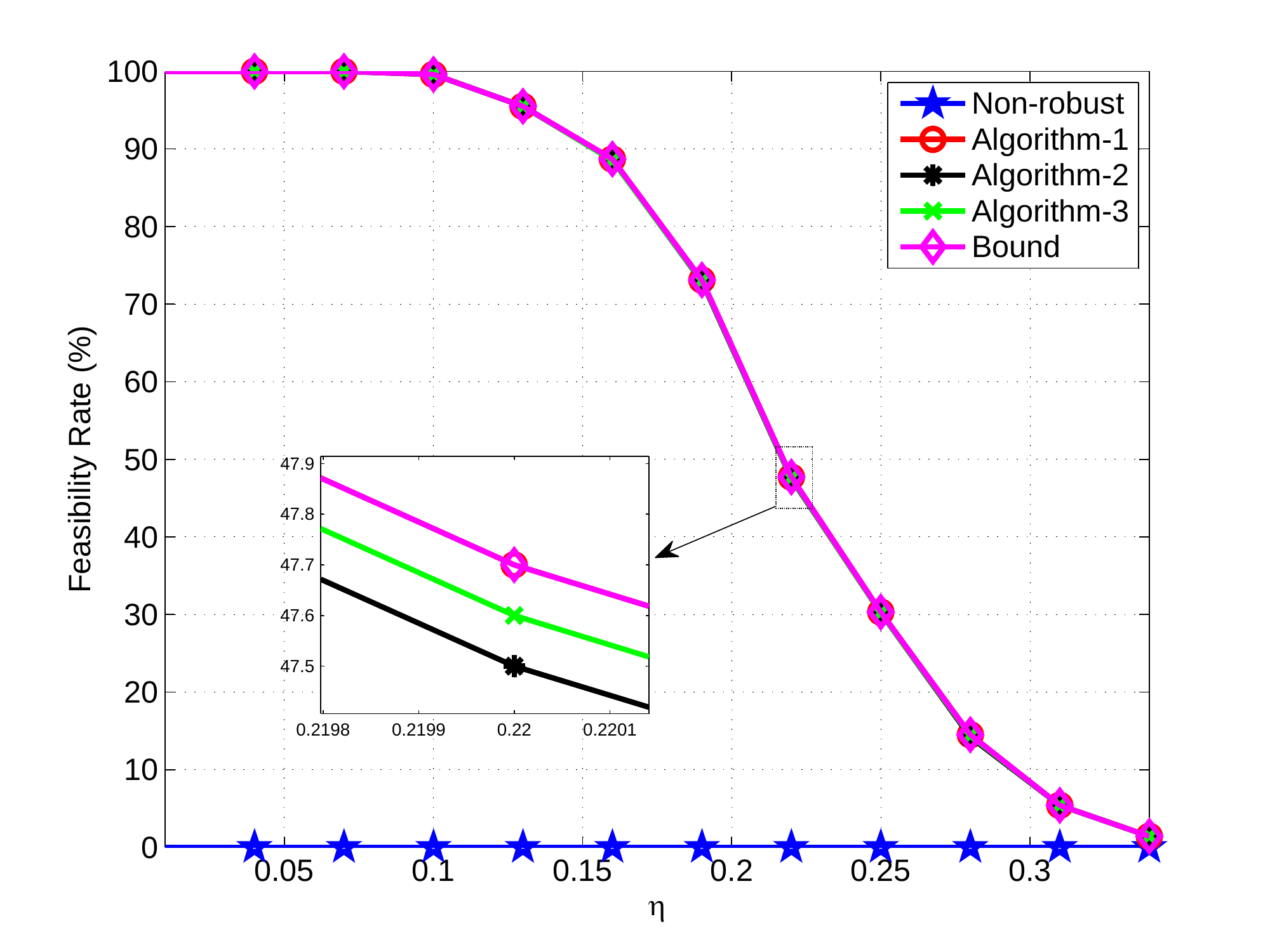}\\
  \caption{Feasibility rate (\%) versus various $\eta$. $N=4$, $\gamma  = 10$ dB, $\psi = 5$ dBm.}\label{feasible_rate_eta_new2}\vspace{-0.8em}
\end{figure}

\begin{figure}[hbtp]
  \setlength{\abovecaptionskip}{-0.2cm} 
  \setlength{\belowcaptionskip}{-0.2cm} 
  \centering
  \includegraphics[width = 0.42\textwidth]{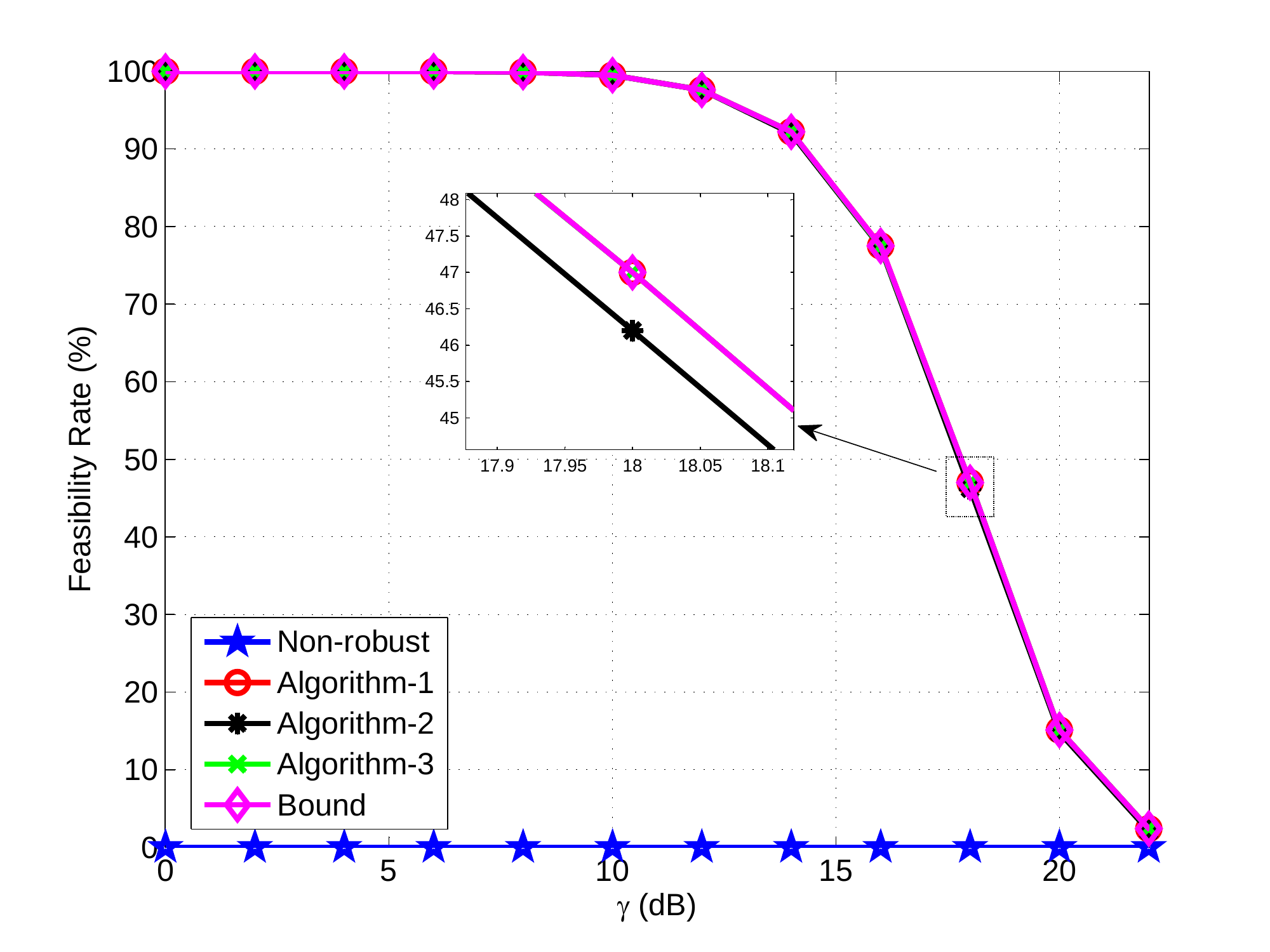}\\
  \caption{Feasibility rate (\%) versus various $\gamma$. $N=4$, $\eta  = 0.1$, $\psi = 5$ dBm.}\label{feasible_rate_gamma_new2}\vspace{-0.8em}
\end{figure}

\emph{1) Feasibility rate:} We first present the feasibility rates of the three robust JBPS design algorithms. In the simulation, a robust design algorithm is considered infeasible for a channel realization if CVX reports an infeasible status or ${\widetilde x_k} \le 0$ in the robust solution recovery method. The feasibility of the non-robust design \cite{BeamformingforMISO} are tested with $100$ channel errors satisfying the NBE model for each channel realization. Fig. \ref{feasible_rate_eta_new2} and Fig. \ref{feasible_rate_gamma_new2} present the simulation results obtained over 1000 channel realizations. One can observe from this figure that the three algorithms exhibit similar (almost identical) feasibility rate compared to the bound.\footnote{The feasibility rate of the bound is equivalent to the feasibility rate of solving problem (\ref{robustSDP}) without rank recovery.} The non-robust method fails to satisfy both the SINR and EH constraints almost all the time under NBE model.

\begin{figure}[hbtp]
  \setlength{\abovecaptionskip}{-0.2cm} 
  \setlength{\belowcaptionskip}{-0.2cm} 
  \centering
  \includegraphics[width = 0.42\textwidth]{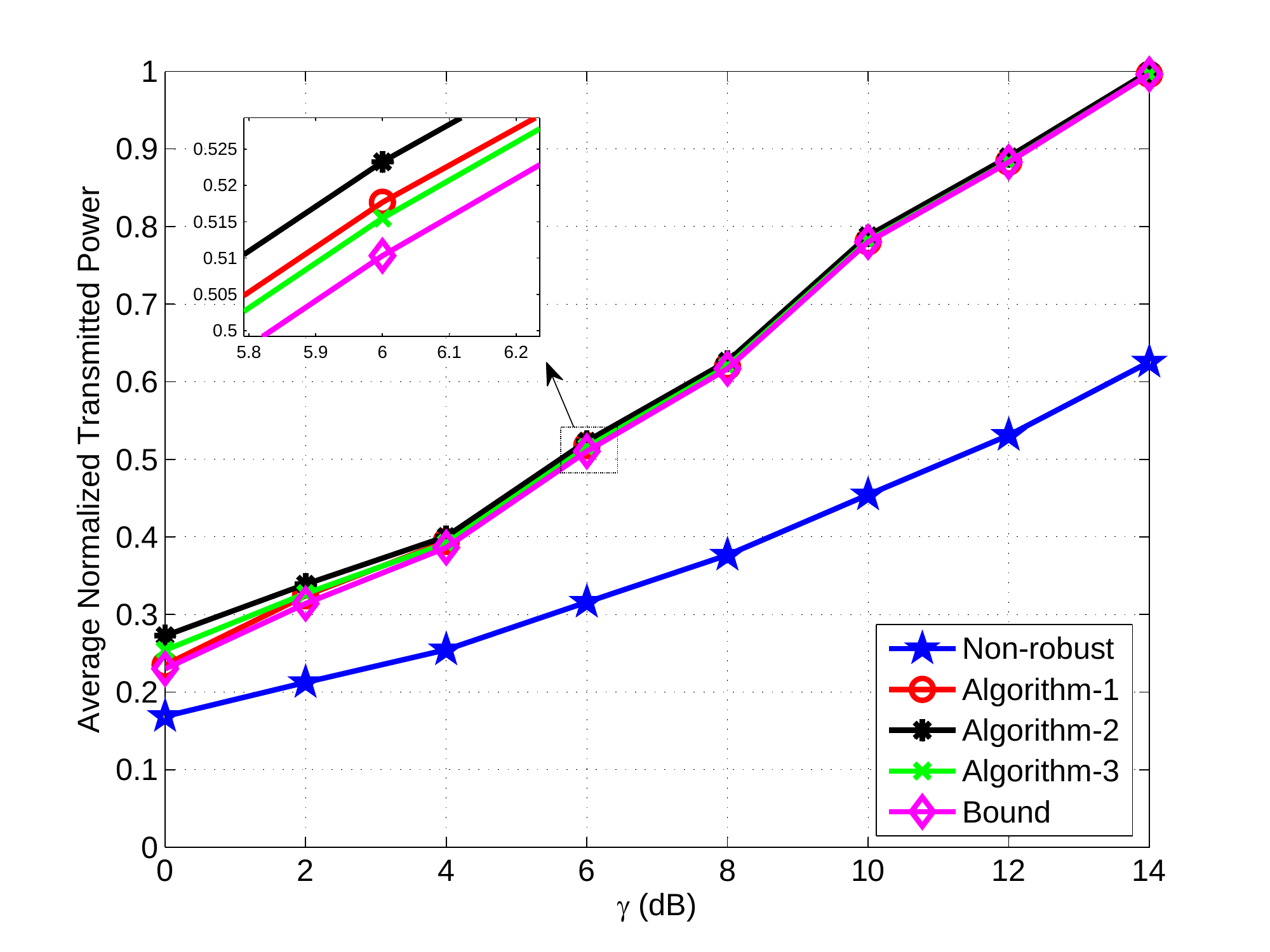}\\
  \caption{Transmission power versus SINR target $\gamma$. $N=4$, $\eta  = 0.1$, $\psi = 5$ dBm.}\label{average_power_gamma_new2}\vspace{-0.8em}
\end{figure}
\begin{figure}[hbtp]
  \setlength{\abovecaptionskip}{-0.2cm} 
  \setlength{\belowcaptionskip}{-0.2cm} 
  \centering
  \includegraphics[width = 0.42\textwidth]{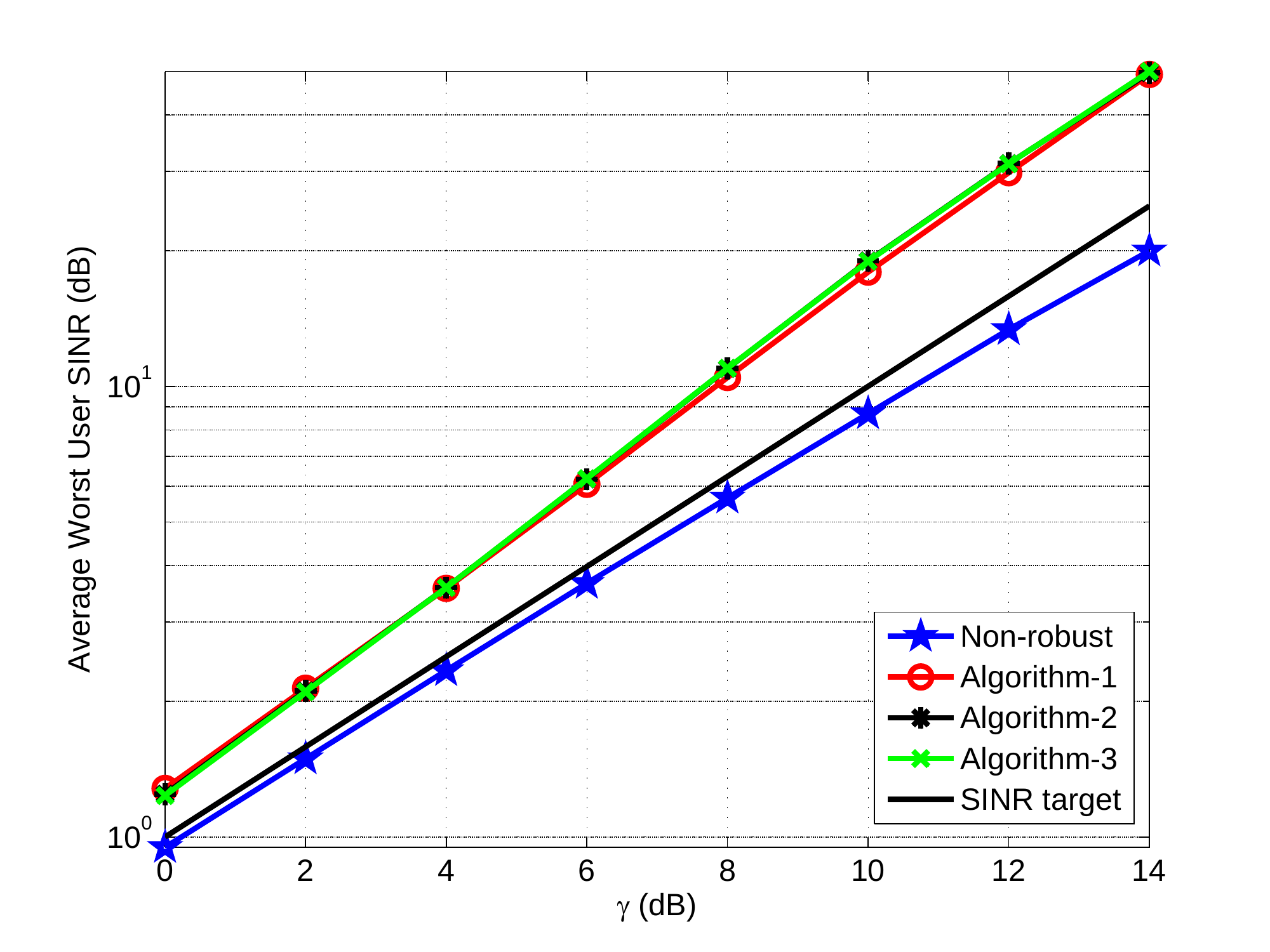}\\
  \caption{Average worst user SINR versus SINR target $\gamma$. $N=4$, $\eta  = 0.1$, $\psi = 5$ dBm.}\label{worst_SINR_gamma_new2}\vspace{-0.8em}
\end{figure}

\emph{2) Transmission power:} We illustrate the performance of the three robust designs in terms of average transmission power over 1000 problem instances. Fig. \ref{average_power_gamma_new2} shows performance comparison among the three robust designs, the transmission power being averaged over problem instances where the robust designs are all feasible. It is observed that, as a price paid for guaranteed worst-case performance, the robust designs require higher average transmission power than the non-robust design. However, \emph{Algorithm-1/2/3} show near-optimal performance compared to the bound. The performance of \emph{Algorithm-3} is slightly better than \emph{Algorithm-1} due to the fact that higher-rank solutions may be returned by solving problem (\ref{robustSDP}). In the low SINR region, almost all the solutions returned by \emph{Algorithm-1} are rank-one (optimal), thus the performance of \emph{Algorithm-3} is slightly inferior to that of \emph{Algorithm-1} since \emph{Algorithm-3} does not converge to the optimal solution in $N^{max}$ iterations.

The average worst user SINR performance of the proposed robust designs is illustrated in Fig. \ref{worst_SINR_gamma_new2} for both the robust and non-robust designs under the same simulation parameters as in Fig. \ref{average_power_gamma_new2}. Clearly, the average achieved worst user SINR of the robust designs are all above the SINR target while the non-robust design fails to satisfy the SINR constraints.

\begin{figure}[hbtp]
  \setlength{\abovecaptionskip}{-0.2cm} 
  \setlength{\belowcaptionskip}{-0.2cm} 
  \centering
  \includegraphics[width = 0.42\textwidth]{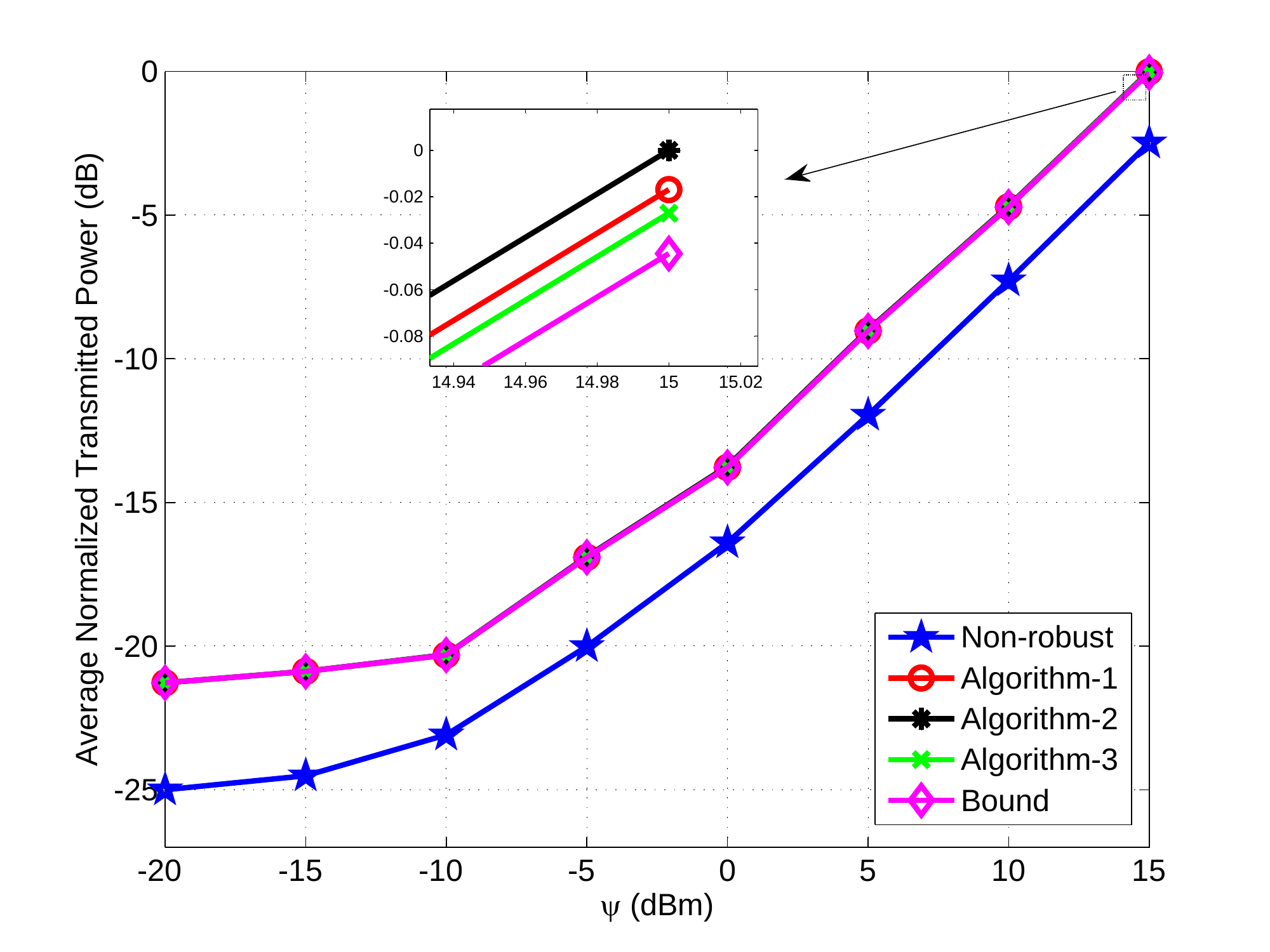}\\
  \caption{Transmission power versus EH target $\psi$. $N=4$, $\eta  = 0.1$, $\gamma = 10$ dB.}\label{average_power_e_new2}\vspace{-0.8em}
\end{figure}

\begin{figure}[hbtp]
  \setlength{\abovecaptionskip}{-0.2cm} 
  \setlength{\belowcaptionskip}{-0.2cm} 
  \centering
  \includegraphics[width = 0.42\textwidth]{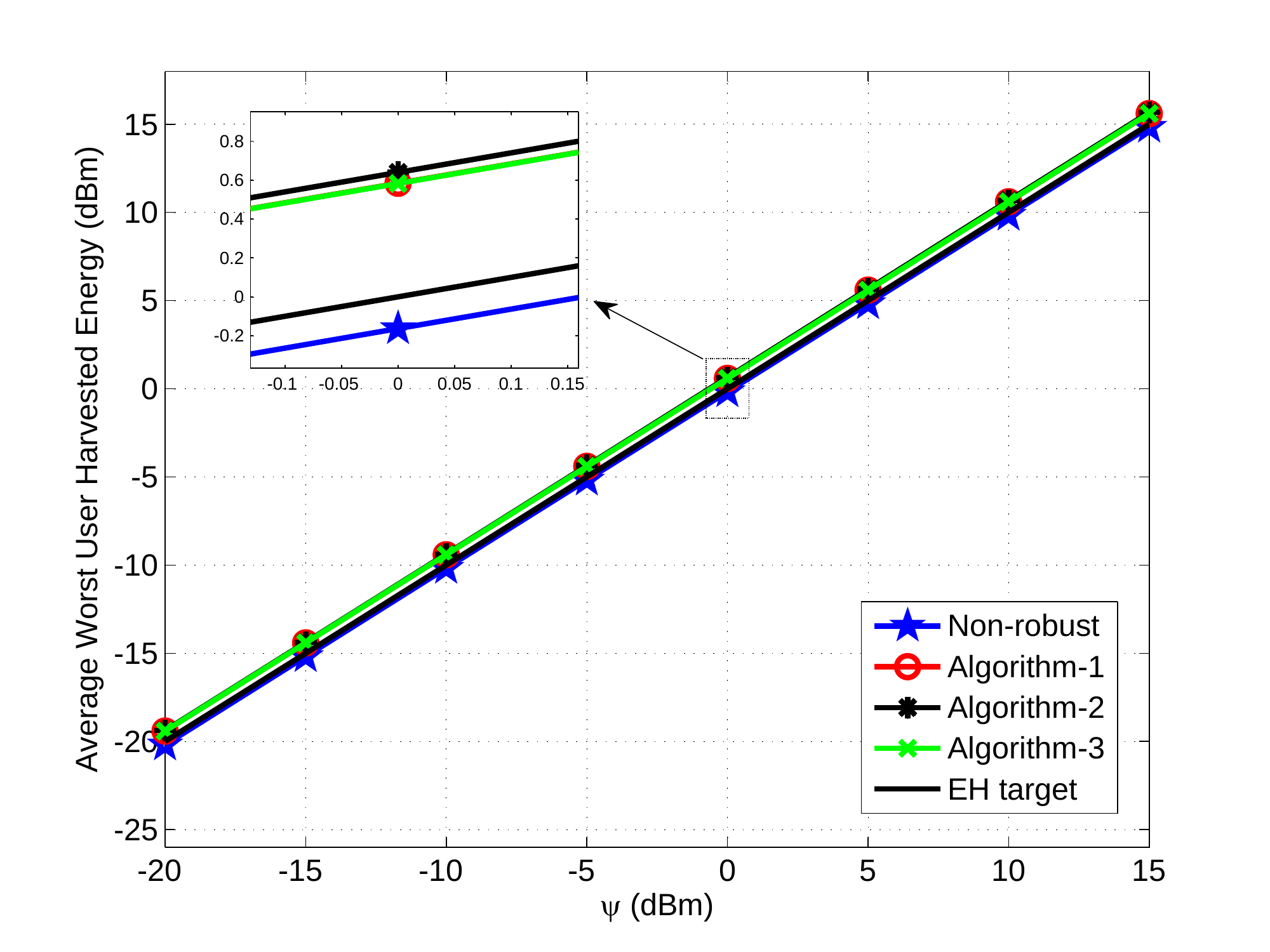}\\
  \caption{Average worst user harvested energy versus EH target $\psi$. $N=4$, $\eta  = 0.1$, $\gamma = 10$ dB.}\label{worst_EH_e_new2}\vspace{-0.8em}
\end{figure}

Fig. \ref{average_power_e_new2} shows performance comparison among the robust designs for various EH constraints. One can see that the robust designs require higher average transmission power than the non-robust design. The best performance is achieved by \emph{Algorithm-3}, followed by \emph{Algorithm-1} and \emph{Algorithm-2}. The average worst
user harvested power of the proposed robust designs is illustrated in Fig. \ref{worst_EH_e_new2}, where the simulation parameters are the same as in Fig. \ref{average_power_e_new2}. Clearly, the average achieved worst user harvested power of the robust designs are all above the EH target while the non-robust design fails to satisfy the EH constraints.

\begin{figure}[hbtp]
  \setlength{\abovecaptionskip}{-0.2cm} 
  \setlength{\belowcaptionskip}{-0.2cm} 
  \centering
  \includegraphics[width = 0.42\textwidth]{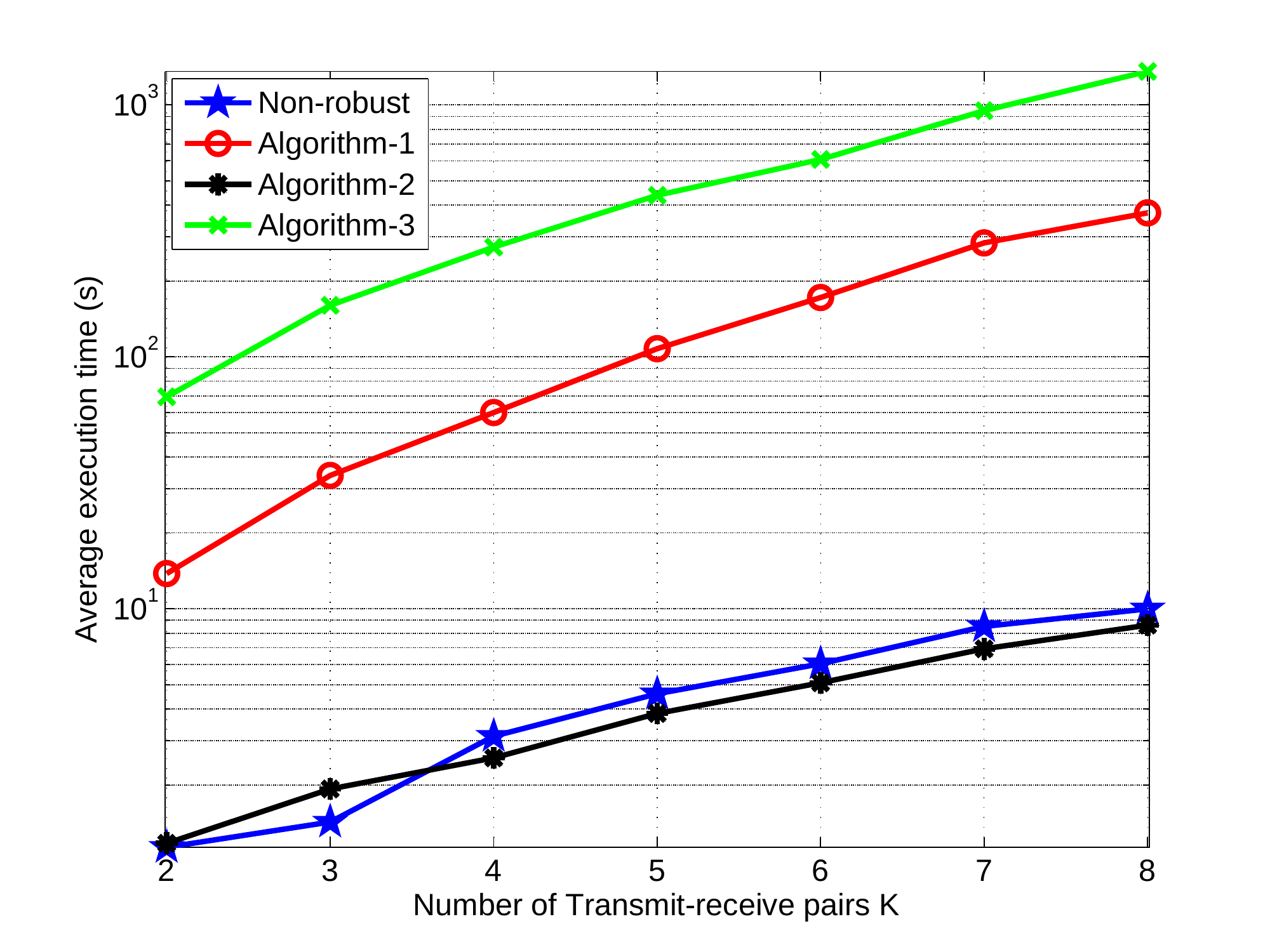}\\
  \caption{Comparison of execution time versus $K$ between the robust and non-robust designs with fixed $N=18$. $\psi = 5$ dBm, $\eta  = 0.1$, $\gamma = 10$ dB.}\label{average_exc_time_K_new2}\vspace{-0.8em}
\end{figure}

%
\emph{3) Execution time:}  We then compare the performance of the robust and non-robust designs in terms of average execution time over $20$ channel realizations. Fig. \ref{average_exc_time_K_new2} demonstrates the execution time (on a logarithm scale) versus the number of transmit-receive pairs with fixed number of transmit antennas $N=18$. It is observed that the time consumed by all four algorithms increases with $K$. However \emph{Algorithm-2} consumes much less time than \emph{Algorithm-1} and \emph{Algorithm-3}, which means that \emph{Algorithm-2} shows great potential for applications with large antenna arrays and large number of transmit-receive pairs. \emph{Algorithm-3} requires the most time as a price for better performance when higher-rank solutions are returned.

\section{Conclusion} \label{section:conclusion}
In this paper, we considered the robust JBPS design problem for multiuser MISO interference channel under a NBE model for the CSI. Three different robust design approaches were proposed to handle the highly non-convex JBPS problem with different performance and complexity. In the first design approach, we proposed to relax the original problem as a SDP problem based on SDR, which provides a lower bound for the robust JBPS problem if a rank-one solution is returned. A rank-one recovery method was provided to obtain a feasible rank-one solution if a high-rank solution is returned. In the second design approach, we formulated the robust JBPS problem as a SOCP problem based on SOCP relaxation and the Cauchy-Schwarz inequality. As compared to the SDR-based algorithm, the SOCP method has lower computational complexity, while achieving a performance very close to the performance bound. We also provided a closed-form robust feasible solution recovery method. In the third design approach, a CCCP-based iterative algorithm was presented to achieve near-optimal performance when a higher-rank solution is returned by the SDR-based algorithm.
It was proved that any limit point of the iterative algorithm is a KKT solution to the robust JBPS problem.
The simulation results showed that the proposed robust transceiver designs have near-optimal performance in the presence of imperfect CSI.

\begin{appendices}

\section{The Proof of Lemma \ref{lamma2}} \label{appendixB}
We introduce the following variables to dualize the corresponding constraints in problem (\ref{iterative_SDP4}) (the notation $A$ : $B$ denotes the constraint $B$ and its dual variable $A$)
\begin{equation}
\begin{array}{l}
{\widetilde {\bf{U}}_k} \buildrel \Delta \over = \left[ {\begin{array}{*{20}{c}}
{{{\overline {\bf{U}} }_k}}&{{{\bf{u}}_k}}\\
{{\bf{u}}_k^{H}}&{{u_k}}
\end{array}} \right]:(\ref{SINRcons1_SDP}),\; {\widetilde {\bf{V}}_{kj}} \buildrel \Delta \over = \left[ {\begin{array}{*{20}{c}}
{{{\overline {\bf{V}} }_{kj}}}&{{{\bf{v}}_{kj}}}\\
{{\bf{v}}_{kj}^{H}}&{{v_{kj}}}
\end{array}} \right]:(\ref{SINRcons2_SDP}),\\
{\widetilde {\bf{X}}_k} \buildrel \Delta \over = \left[ {\begin{array}{*{20}{c}}
{{{\overline {\bf{X}} }_k}}&{{{\bf{x}}_k}}\\
{{\bf{x}}_k^{H}}&{{x_k}}
\end{array}} \right]:(\ref{EHcons1_SDP}),\; {\widetilde {\bf{Y}}_{kj}} \buildrel \Delta \over = \left[ {\begin{array}{*{20}{c}}
{{{\overline {\bf{Y}} }_{kj}}}&{{{\bf{y}}_{kj}}}\\
{{\bf{y}}_{kj}^{H}}&{{y_{kj}}}
\end{array}} \right]:(\ref{EHcons2_SDP}),\\
{a_{kj}}:{\lambda _{kj}}\geq 0,\;{b_{kj}}:{\mu _{kj}}\geq0,\\
{{c}_{kj}}:{p_{kj}}\geq0,\, {{d}_{kj}}:{q_{kj}}\geq0,\; j \neq k,\\
e:(\ref{power_constraint}),\;  \widetilde \alpha_k : \alpha_k \geq 1,\;\widetilde \beta_k: \beta_k \geq 1,\;h_k : (\ref{lamma2_2}),\\
{{\bf{Z}}_k}:{{\bf{F}}_k} \succeq {\bf{0}},\;{\bf{A}}_k : (\ref{lamma2_3}).
\end{array}
\end{equation}
Then the dual problem of problem (\ref{iterative_SDP4}) can be expressed as
\begin{equation} \label{Dual_problem}
\mathop {\max }\limits_{\left\{ \begin{array}{l}{\scriptstyle {{\bf{Z}}_k}\succeq{\bf{0}},{{\widetilde {\bf{U}}}_k}\succeq{\bf{0}},{{\widetilde {\bf{V}}}_{kj}}\succeq{\bf{0}},{{\widetilde {\bf{X}}}_k}\succeq{\bf{0}},}
\\
\scriptstyle { {{{\widetilde {\bf{Y}}}_{kj}}\succeq{\bf{0}}, {a_{kj}} \ge 0,{b_{kj}} \ge 0,{c_{kj}} \ge 0,{d_{kj}} \ge 0}}\\
\scriptstyle { {{\bf{Z}}_k}\succeq{\bf{0}},{e} \ge 0,{\widetilde \alpha_k} \ge 0,{\widetilde \beta_k} \ge 0,{h_k} \ge 0}\\
\end{array} \right\}} \mathop {\inf }\limits_{\left\{ \begin{array}{l}{\scriptstyle{{{\bf{F}}_k},{\lambda _{kj}},{\mu _{kj}},}} \\
 \scriptstyle{ {{\bf{f}}_k}, \alpha_k, \beta_k , t} \\
 \scriptstyle{{p_{kj}},{q_{kj}}} \end{array} \right\}}
  \mathcal{L}
\end{equation}
where $\mathcal{L}$ can be expressed as (\ref{appendixB-objectivefunc}), shown at the top of the next page.
\begin{figure*}
\vspace{-1.2em}
\begin{equation}\label{appendixB-objectivefunc}
\begin{array}{l}
\mathcal{L} = {t^2} + \sum\limits_{k = 1}^K {\textrm{Tr}\{ { [ \begin{array}{l}
{\bf{A}}_k - \frac{1}{{{\gamma _k}}}{\overline{\overline {\bf{U}}} _k} + \sum\limits_{j \ne k}^K {\overline{\overline {\bf{V}}} _{jk}}
 - {\overline{\overline {\bf{X}}} _k} - \sum\limits_{j \ne k}^K {{\overline{\overline {\bf{Y}}} _{jk}} - {{\bf{Z}}_k}}
\end{array}]{{\bf{F}}_k}} \}} + \sum\limits_{k = 1}^K { {{u_k}\sigma _k^2  - {x_k}\sigma _k^2  }}\\
+\sum\limits_{k = 1}^K {( {-{a_{kk}} - \textrm{Tr}\left( {{{\overline {\bf{U}} }_k}} \right) + {u_k}\eta _{kk}^2} ){\lambda _{kk}}}  + \sum\limits_{k = 1}^K {\sum\limits_{j \ne k}^K {( {-{a_{kj}} - \textrm{Tr}\left( {{{\overline {\bf{V}} }_{kj}}} \right) + {v_{kj}}\eta _{kj}^2} ){\lambda _{kj}}} } + \sum\limits_{k = 1}^K {\sum\limits_{j \ne k}^K {( { - {c_{kj}} + {u_k} - {v_{kj}}} ){p_{kj}}} } \\
+\sum\limits_{k = 1}^K {( {-{{{b}}_{kk}} - \textrm{Tr}\left( {{{\overline {\bf{X}} }_k}} \right) + {x_k}\eta _{kk}^2} ){\mu _{kk}}}  + \sum\limits_{k = 1}^K {\sum\limits_{j \ne k}^K {( {-{{{b}}_{kj}} - \textrm{Tr}\left( {{{\overline {\bf{Y}} }_{kj}}} \right) + {y_{kj}}\eta _{kj}^2} ){\mu _{kj}}} } + \sum\limits_{k = 1}^K {\sum\limits_{j \ne k}^K {( { - {d_{kj}} - {x_k} + {y_{kj}}} ){q_{kj}}} },\\
+\sum\limits_{k = 1}^K {({u_k}\omega _k^2 - {\widetilde \alpha_k}){\alpha _k} + \left( {{x_k}\frac{{{\psi _k}}}{{{\xi _k}}} - {\widetilde \beta_k}} \right){\beta _k} + {\widetilde \alpha_k} + {\widetilde \beta_k} + {h_k}({\textrm{invp}}({\alpha _k}) + {\textrm{invp}}({\beta _k}) - 1)} \\
+\sum\limits_{k = 1}^K {{\textrm{Tr}}({{\bf{A}}_k}({\bf{f}}_k^i{{({{\bf{f}}_k} - {\bf{f}}_k^i)}^H} + ({{\bf{f}}_k} - {\bf{f}}_k^i){\bf{f}}_k^{iH} + {\bf{f}}_k^i{\bf{f}}_k^{iH}))-e\left( {\sqrt {\sum\limits_{k = 1}^K {\textrm{Tr}({{\bf{f}}_k}{\bf{f}}_k^H)} }  - t} \right)}, \\
{\overline{\overline {\bf{U}}} _k} = {{{\overline {\bf{U}} }_k} + {{\bf{u}}_k}\widehat {\bf{h}}_{kk}^H + \widehat {\bf{h}}_{kk}^{}{\bf{u}}_k^H + {u_k}\widehat {\bf{h}}_{kk}^{}\widehat {\bf{h}}_{kk}^H},\; {\overline{\overline {\bf{V}}} _{jk}} = {{{\overline {\bf{V}} }_{jk}} + {{\bf{v}}_{jk}}\widehat {\bf{h}}_{jk}^H + \widehat {\bf{h}}_{jk}^{}{\bf{v}}_{jk}^H + {v_{jk}}\widehat {\bf{h}}_{jk}^{}\widehat {\bf{h}}_{jk}^H},\\
{\overline{\overline {\bf{X}}} _k} = {{{\overline {\bf{X}} }_k} + {{\bf{x}}_k}\widehat {\bf{h}}_{kk}^H + \widehat {\bf{h}}_{kk}^{}{\bf{x}}_k^H + {x_k}\widehat {\bf{h}}_{kk}^{}\widehat {\bf{h}}_{kk}^H},\; {\overline{\overline {\bf{Y}}} _{jk}} = {{{\overline {\bf{Y}} }_{jk}} + {{\bf{y}}_{jk}}\widehat {\bf{h}}_{jk}^H + \widehat {\bf{h}}_{jk}^{}{\bf{y}}_{jk}^H + {y_{jk}}\widehat {\bf{h}}_{jk}^{}\widehat {\bf{h}}_{jk}^H}.
\end{array}
\end{equation}
\hrulefill
\end{figure*}
Assume $\| {{{\widehat {\bf{h}}}_{kk}}\widehat {\bf{h}}_{kk}^H} \| = h_{kj} \| {{{\widehat {\bf{h}}}_{kj}}\widehat {\bf{h}}_{kj}^H} \|, \; h_{kj}>0$, then if we choose ${\widetilde {\bf{U}}_k}$, ${\widetilde {\bf{V}}_{kj}}$, ${\widetilde {\bf{X}}_k}$ and ${\widetilde {\bf{Y}}_{kj}}$ as in (\ref{choose}),
\begin{figure*}
\vspace{-0.6em}
\begin{equation} \label{choose}
\begin{array}{l}
{\widetilde {\bf{U}}_k} \buildrel \Delta \over = \left[ {\begin{array}{*{20}{c}}
{{{\overline {\bf{U}} }_k}}&{{{\bf{u}}_k}}\\
{{\bf{u}}_k^H}&{{u_k}}
\end{array}} \right] = {\textrm{diag}}\left\{ {\min \left\{ {\frac{{{\gamma _k}}}{{{a_1}}},\frac{{{\gamma _k}\eta _{kk}^2}}{{2{a_1}{N_k}\left\| {{{\widehat {\bf{h}}}_{kk}}\widehat {\bf{h}}_{kk}^H} \right\|}}} \right\}{{\bf{I}}_{{N_k}}},\frac{{{\gamma _k}}}{{{a_1}\left\| {{{\widehat {\bf{h}}}_{kk}}\widehat {\bf{h}}_{kk}^H} \right\|}}} \right\} \succ {\bf{0}},\\
{\widetilde {\bf{V}}_{kj}} \buildrel \Delta \over = \left[ {\begin{array}{*{20}{c}}
{{{\overline {\bf{V}} }_{kj}}}&{{{\bf{v}}_{kj}}}\\
{{\bf{v}}_{kj}^H}&{{v_{kj}}}
\end{array}} \right] = {\textrm{diag}}\left\{ {\min \left\{ {\frac{1}{{{a_2}}},\frac{{\eta _{kj}^2}}{{2{a_2}{N_k}\left\| {{{\widehat {\bf{h}}}_{kj}}\widehat {\bf{h}}_{kj}^H} \right\|}}} \right\}{{\bf{I}}_{{N_k}}},\frac{1}{{{a_2}\left\| {{{\widehat {\bf{h}}}_{kj}}\widehat {\bf{h}}_{kj}^H} \right\|}}} \right\} \succ {\bf{0}},\\
{\widetilde {\bf{X}}_k} \buildrel \Delta \over = \left[ {\begin{array}{*{20}{c}}
{{{\overline {\bf{X}} }_k}}&{{{\bf{x}}_k}}\\
{{\bf{x}}_k^H}&{{x_k}}
\end{array}} \right] = {\textrm{diag}}\left\{ {\min \left\{ {\frac{1}{{{a_3}(K + 1)}},\frac{{\eta _{kk}^2}}{{2{a_3}{N_k}(K + 1)\left\| {{{\widehat {\bf{h}}}_{kk}}\widehat {\bf{h}}_{kk}^H} \right\|}}} \right\}{{\bf{I}}_{{N_k}}},\frac{1}{{{a_3}(K + 1)\left\| {{{\widehat {\bf{h}}}_{kk}}\widehat {\bf{h}}_{kk}^H} \right\|}}} \right\} \succ {\bf{0}},\\
{\widetilde {\bf{Y}}_{kj}} \buildrel \Delta \over = \left[ {\begin{array}{*{20}{c}}
{{{\overline {\bf{Y}} }_{kj}}}&{{{\bf{y}}_{kj}}}\\
{{\bf{y}}_{kj}^H}&{{y_{kj}}}
\end{array}} \right] = {\textrm{diag}}\left\{ {\min \left\{ {\frac{1}{{{a_1}(K - 1)}},\frac{{\eta _{kj}^2}}{{2{a_1}(K - 1){N_k}\left\| {{{\widehat {\bf{h}}}_{kj}}\widehat {\bf{h}}_{kj}^H} \right\|}}} \right\}{{\bf{I}}_{{N_k}}},\frac{1}{{{a_1}(K - 1)\left\| {{{\widehat {\bf{h}}}_{kj}}\widehat {\bf{h}}_{kj}^H} \right\|}}} \right\} \succ {\bf{0}},
\end{array}
\end{equation}
\hrulefill
\end{figure*}
$\forall k \in \mathcal{K}$, $j \neq k$, with $a_2 > a_1 \frac {h_{kj}}{\gamma_k}$, $a_3 > \frac {a_1} {h_{kj}}$ and $1> \frac{4}{a_1} + \frac{2}{a_3(K+1)}, \;a_1,a_2,a_3 \in \mathbb{R}_+$, and choose ${{\bf{A}}_k} = {\bf{I}}$, $e$ and $h_k$ as any positive real number, we have
\begin{equation}
\begin{array}{l}
{a_{kk}} \buildrel \Delta \over = {u_k}\eta _{kk}^2 - {\textrm{Tr}}( {{{\overline {\bf{U}} }_k}} )\\
= \frac{{{\gamma _k}\eta _{kk}^2}}{{{a_1}\| {{{\widehat {\bf{h}}}_{kk}}\widehat {\bf{h}}_{kk}^H} \|}} - \min \{ {\frac{{{\gamma _k}}}{{{a_1}}},\frac{{{\gamma _k}\eta _{kk}^2}}{{2{a_1}{N_k}\| {{{\widehat {\bf{h}}}_{kk}}\widehat {\bf{h}}_{kk}^H} \|}}} \}{\textrm{Tr}}( {{{\bf{I}}_{{N_k}}}} )\\
\ge \frac{{{\gamma _k}\eta _{kk}^2}}{{{a_1}\| {{{\widehat {\bf{h}}}_{kk}}\widehat {\bf{h}}_{kk}^H} \|}} - \frac{{{\gamma _k}\eta _{kk}^2{N_k}}}{{2{a_1}{N_k}\| {{{\widehat {\bf{h}}}_{kk}}\widehat {\bf{h}}_{kk}^H} \|}} = \frac{{{\gamma _k}\eta _{kk}^2}}{{2{a_1}\| {{{\widehat {\bf{h}}}_{kk}}\widehat {\bf{h}}_{kk}^H} \|}} > 0,
\end{array}
\end{equation}
\begin{equation}
\begin{array}{l}
{a_{kj}} \buildrel \Delta \over = {v_{kj}}\eta _{kj}^2 - {\textrm{Tr}}( {{{\overline {\bf{V}} }_{kj}}} )\\
= \frac{{\eta _{kj}^2}}{{{a_2}\| {{{\widehat {\bf{h}}}_{kj}}\widehat {\bf{h}}_{kj}^H} \|}} - \min \{ {\frac{1}{{{a_2}}},\frac{{\eta _{kj}^2}}{{2{a_2}{N_k}\| {{{\widehat {\bf{h}}}_{kj}}\widehat {\bf{h}}_{kj}^H} \|}}} \}{\textrm{Tr}}( {{{\bf{I}}_{{N_k}}}} )\\
\ge \frac{{\eta _{kj}^2}}{{{a_2}\| {{{\widehat {\bf{h}}}_{kj}}\widehat {\bf{h}}_{kj}^H} \|}} - \frac{{\eta _{kj}^2{N_k}}}{{2{a_2}{N_k}\| {{{\widehat {\bf{h}}}_{kj}}\widehat {\bf{h}}_{kj}^H} \|}} = \frac{{\eta _{kj}^2}}{{2{a_2}\| {{{\widehat {\bf{h}}}_{kj}}\widehat {\bf{h}}_{kj}^H} \|}} > 0.
\end{array}
\end{equation}
Following the same derivation, we can obtain ${b_{kk}}  \ge 0$ and ${b_{kj}} \ge 0$. Furthermore, we have
\begin{equation}
\begin{array}{l}
{c_{kj}} \buildrel \Delta \over = {u_k} - {v_{kj}} = \frac{{{\gamma _k}}}{{{a_1}\left\| {{{\widehat {\bf{h}}}_{kk}}\widehat {\bf{h}}_{kk}^H} \right\|}} - \frac{1}{{{a_2}\left\| {{{\widehat {\bf{h}}}_{kj}}\widehat {\bf{h}}_{kj}^H} \right\|}}\\
> \frac{{{\gamma _k}}}{{{a_1}\left\| {{{\widehat {\bf{h}}}_{kk}}\widehat {\bf{h}}_{kk}^H} \right\|}} - \frac{{{\gamma _k}}}{{{a_1}{h_{kj}}\left\| {{{\widehat {\bf{h}}}_{kj}}\widehat {\bf{h}}_{kj}^H} \right\|}} = 0,
\end{array}
\end{equation}
\begin{equation}
\begin{array}{l}
{d_{kj}} \buildrel \Delta \over = {y_{kj}} - {x_k}\\
= \frac{1}{{{a_1}(K - 1)\| {{{\widehat {\bf{h}}}_{kj}}\widehat {\bf{h}}_{kj}^H} \|}} - \frac{1}{{{a_3}(K + 1)\| {{{\widehat {\bf{h}}}_{kk}}\widehat {\bf{h}}_{kk}^H} \|}}\\
> \frac{1}{{{a_1}(K - 1)\| {{{\widehat {\bf{h}}}_{kj}}\widehat {\bf{h}}_{kj}^H} \|}} - \frac{1}{{{a_1}(K + 1)\| {{{\widehat {\bf{h}}}_{kj}}\widehat {\bf{h}}_{kj}^H} \|}} > 0,\\
\end{array}
\end{equation}
\begin{equation}
{\widetilde \alpha_k} \buildrel \Delta \over = {u_k}\omega _k^2,\; {\widetilde \beta_k} \buildrel \Delta \over = {x_k}\frac{{{\psi _k}}}{{{\xi _k}}},
\end{equation}
and ${{\bf{Z}}_k}$ can be expressed as (\ref{Z_k}).
\begin{figure*}
\begin{equation} \label{Z_k}
\begin{array}{l}
{{\bf{Z}}_k} \buildrel \Delta \over = {\bf{I}} - \frac{1}{{{\gamma _k}}}{\overline {\overline {\bf{U}} } _k} + \sum\limits_{j \ne k}^K {{{\overline {\overline {\bf{V}} } }_{jk}}}  - {\overline {\overline {\bf{X}} } _k} - \sum\limits_{j \ne k}^K {{{\overline {\overline {\bf{Y}} } }_{jk}}} \succeq {\bf{I}} - \frac{1}{{{\gamma _k}}}{\overline {\overline {\bf{U}} } _k} - {\overline {\overline {\bf{X}} } _k} - \sum\limits_{j \ne k}^K {{{\overline {\overline {\bf{Y}} } }_{jk}}}  = {\bf{I}} - \frac{1}{{{\gamma _k}}}\min \{ \frac{{{\gamma _k}}}{{{a_1}}},\frac{{{\gamma _k}\eta _{kk}^2}}{{2{a_1}{N_k}{{\widehat {\bf{h}}}_{kk}}\widehat {\bf{h}}_{kk}^H}}\} {{\bf{I}}_{{N_k}}}\\
 - \min \{ \frac{1}{{{a_3}(K + 1)}},\frac{{\eta _{kk}^2}}{{2{a_3}{N_k}(K + 1){{\widehat {\bf{h}}}_{kk}}\widehat {\bf{h}}_{kk}^H}}\} {{\bf{I}}_{{N_k}}} - \sum\limits_{j = 1,j \ne k}^K {\min } \{ \frac{1}{{{a_1}(K - 1)}},\frac{{\eta _{jk}^2}}{{2{a_1}(K - 1){N_k}{{\widehat {\bf{h}}}_{jk}}\widehat {\bf{h}}_{jk}^H}}\} {{\bf{I}}_{{N_k}}} - \frac{{\widehat {\bf{h}}_{kk}^{}\widehat {\bf{h}}_{kk}^H}}{{{a_3}(K + 1){{\widehat {\bf{h}}}_{kk}}\widehat {\bf{h}}_{kk}^H}}\\
 - \frac{{{{\widehat {\bf{h}}}_{kk}}\widehat {\bf{h}}_{kk}^H}}{{{a_1}{{\widehat {\bf{h}}}_{kk}}\widehat {\bf{h}}_{kk}^H}} - \sum\limits_{j = 1,j \ne k}^K {\frac{{{{\widehat {\bf{h}}}_{jk}}\widehat {\bf{h}}_{jk}^H}}{{{a_1}(K - 1){{\widehat {\bf{h}}}_{jk}}\widehat {\bf{h}}_{jk}^H}}}\succeq {\bf{I}} - \frac{1}{{{\gamma _k}}}(\frac{{{\gamma _k}}}{{{a_1}}}{\bf{I}} + \frac{{{\gamma _k}{{\widehat {\bf{h}}}_{kk}}\widehat {\bf{h}}_{kk}^H}}{{{a_1}{{\widehat {\bf{h}}}_{kk}}\widehat {\bf{h}}_{kk}^H}}) - (\frac{1}{{{a_3}(K + 1)}}{\bf{I}} + \frac{{\widehat {\bf{h}}_{kk}^{}\widehat {\bf{h}}_{kk}^H}}{{{a_3}(K + 1){{\widehat {\bf{h}}}_{kk}}\widehat {\bf{h}}_{kk}^H}})\\
 - \sum\limits_{j = 1,j \ne k}^K {(\frac{1}{{{a_1}(K - 1)}}{{\bf{I}}_{{N_k}}} + \frac{{{{\widehat {\bf{h}}}_{jk}}\widehat {\bf{h}}_{jk}^H}}{{{a_1}(K - 1){{\widehat {\bf{h}}}_{jk}}\widehat {\bf{h}}_{jk}^H}})}\succeq (1 - \frac{4}{{{a_1}}} - \frac{2}{{{a_3}(K + 1)}}){\bf{I}} \succ {\bf{0}},
\end{array}
\end{equation}
\hrulefill
\end{figure*}
which implies that $\{{\widetilde {\bf{U}}_k}\}$, $\{{\widetilde {\bf{V}}_{kj}}\}$, $\{{\widetilde {\bf{X}}_k}\}$ and $\{{\widetilde {\bf{Y}}_{kj}}\}$ define a strictly feasible point of the dual problem and the resulting dual problem is a bounded problem.
Thus, the dual problem is always strictly feasible. Together with the fact that problem (\ref{iterative_SDP4}) is feasible, we can see that Slater's condition always holds for the dual problem (\ref{Dual_problem}). Hence, problem (\ref{iterative_SDP4}) can attain its minimum and strong duality holds \cite{ConvexOptimization}. This completes the proof.
\section{The proof of Lemma \ref{lamma3} } \label{appendixC}
In this appendix, we extend the convergence proof of \cite{2014arXiv1403.3196S} to the case of matrix function. We denote problem (\ref{iterative_SDP4}) by $\mathcal{P}({{\bf{f}}_k^i})$. Define ${\bf{R}}({{\bf{f}}_k},{{\bf{F}}_k}) \buildrel \Delta \over = {{\bf{f}}_k}{\bf{f}}_k^H - {{\bf{F}}_k}$ and $\overline {\bf{R}} ({{\bf{f}}_k},{\bf{f}}_k^i,{{\bf{F}}_k}) \buildrel \Delta \over = {\bf{f}}_k^i{({{\bf{f}}_k} - {\bf{f}}_k^i)^H} + ({{\bf{f}}_k} - {\bf{f}}_k^i){\bf{f}}_k^{iH} + {\bf{f}}_k^i{\bf{f}}_k^{iH} - {{\bf{F}}_k}$. It follows that $\overline {\bf{R}} ({{\bf{f}}_k},{{\bf{f}}_k},{{\bf{F}}_k}) = {\bf{R}}({{\bf{f}}_k},{{\bf{F}}_k})$. In the following, we complete the proof through three steps.

\textbf{In the first step}, we show that each $\{{\bf{f}}_k^i\},i = 1,2, \ldots $ is feasible to problem (\ref{iterative_SDP3}). To this end, it suffices to show that $\{{\bf{f}}_k^{i + 1}\}$ is a feasible solution of problem (\ref{iterative_SDP3}), provided that $\{{\bf{f}}_k^{i}\}$ is feasible. Assume $\{{\bf{f}}_k^{i}\}$ is feasible to problem (\ref{iterative_SDP3}), thus we have $\overline {\bf{R}} ({\bf{f}}_k^i,{\bf{f}}_k^i,{{\bf{F}}_k}) = {\bf{R}}({\bf{f}}_k^i,{{\bf{F}}_k}) = {\bf{f}}_k^i{\bf{f}}_k^{iH} - {{\bf{F}}_k}\succeq{\bf{0}}$. It follows that there must exist $\{{\bf{f}}_k^{i + 1}\}$ that is feasible to problem $\mathcal{P}({\bf{f}}_k^i)$, that is, $\{{\bf{f}}_k^{i + 1}\}$ is such that $\overline {\bf{R}} ({\bf{f}}_k^{i + 1},{\bf{f}}_k^i,{{\bf{F}}_k})\succeq{\bf{0}}$. It follows that
\begin{equation}
\begin{array}{l}
{\bf{R}}({\bf{f}}_k^{i + 1},{{\bf{F}}_k}) = {\bf{f}}_k^{i + 1}{\bf{f}}_k^{(i + 1)H} - {{\bf{F}}_k}\succeq\overline {\bf{R}} ({\bf{f}}_k^{i + 1},{\bf{f}}_k^i,{{\bf{F}}_k})\\
= {\bf{f}}_k^i{({\bf{f}}_k^{i + 1} - {\bf{f}}_k^i)^H} + ({\bf{f}}_k^{i + 1} - {\bf{f}}_k^i){\bf{f}}_k^{iH} + {\bf{f}}_k^i{\bf{f}}_k^{iH} - {{\bf{F}}_k}\succeq{\bf{0}},
\end{array}
\end{equation}
where the first matrix inequality comes from (\ref{CCCP_5}).
This implies that $\{{\bf{f}}_k^{i+1}\}$ is feasible to problem (\ref{iterative_SDP3}). This completes the first step.

\textbf{In the second step}, we show that the objective value sequence $\{P({\bf{f}}_k^i)\}$ monotonically decreases as the iteration index $i$ increases. We denote the optimal solution to $\mathcal{P}({{\bf{f}}_k^i})$ in the $i$th iteration by $\{ {\bf{f}}_k^i,{\bf{F}}_k^i,{\rho ^i}\} $. According to the first step, $\{ {\bf{f}}_k^i,{\bf{F}}_k^i,{\rho ^i}\} $ is a feasible solution to $\mathcal{P}({{\bf{f}}_k^i})$. Moreover, in the $(i+1)$th iteration, the solution $\{ {\bf{f}}_k^{i+1},{\bf{F}}_k^{i+1},{\rho ^{i+1}}\} $ is the optimal solution to $\mathcal{P}({{\bf{f}}_k^i})$. Thus, we have $P({\bf{f}}_k^{i+1})\leq P({\bf{f}}_k^i)$, implying
the monotonic convergence of $\{P({\bf{f}}_k^i)\}$ since $P({\bf{f}}_k)$ is bounded below and $P({\bf{f}}_k^i)$ is convergent. This completes the second step.

\textbf{In the third step}, we prove that any limit point $\{ {\bf{f}}_k^*,{\bf{F}}_k^*,{\rho _k^*}\} $ of the iterates $\{ {\bf{f}}_k^i,{\bf{F}}_k^i,{\rho_k ^i}\} $ is a KKT point of problem (\ref{SDRproblem}).

Let $\mathbb{S}({\bf{f}}_k^i)$ and $\mathbb{C}({\bf{f}}_k^i)$ denote the solution set and constraint set of problem $\mathcal{P}({{\bf{f}}_k^i})$. We first prove ${\bf{f}}_k^* \in \mathbb{S}({\bf{f}}_k^*)$. Since $\{ {\bf{f}}_k^*\} $ is a limit point of $\{ {\bf{f}}_k^i\} $, there must exist a convergent subsequence $\{ {\bf{f}}_k^{i_j}\} $ such that $\textrm{li}{\textrm{m}_{j \to \infty }}{\bf{f}}_k^{i_j} = {\bf{f}}_k^*$. Since the objective function $P({\bf{f}}_k)$ of $\mathcal{P}({{\bf{f}}_k^i})$ is strictly convex in $\{ {{\bf{f}}_k} \in {\mathbb{C}^{{N_k} \times 1}}\} $, the point $\{ {{\bf{f}}_k^{i+1}} \} $ is unique \cite{ConvexOptimization}. Hence, the entries of the two sequences, $\{P({\bf{f}}_k^i)\}$ and $\{ {{\bf{f}}_k^{i}} \} $, have a one-to-one correspondence. By restricting to a subsequence, we can assume that $\{ {\bf{f}}_k^{i_j+1}\} $ converges to a limit point ${\bf{f}}_k^{**}$.

Define the constraint set $\mathbb{C}{_{\succ}}({\widetilde {\bf{f}}_k})  \buildrel \Delta \over = \{ {{{\bf{f}}_k}| {\overline {\bf{R}} ({{\bf{f}}_k},{{\widetilde {\bf{f}}}_k},{{\bf{F}}_k})\succ{\bf{0}}} } \}$ and $\mathbb{C}{_{\succeq}}({\widetilde {\bf{f}}_k})  \buildrel \Delta \over = \{ {{{\bf{f}}_k}| {\overline {\bf{R}} ({{\bf{f}}_k},{{\widetilde {\bf{f}}}_k},{{\bf{F}}_k})\succeq{\bf{0}}}} \}$. It follows that $\mathbb{C}{_{\succ}}({\widetilde {\bf{f}}_k}) \subset \mathbb{C}{_{\succeq}}({\widetilde {\bf{f}}_k}),\; \forall {\widetilde {\bf{f}}_k}$. Let us consider the set $\mathbb{C}{_{\succ}}({ {\bf{f}}_k^*})$. Since ${\overline {\bf{R}} ({{\bf{f}}_k},{{\widetilde {\bf{f}}}_k},{{\bf{F}}_k})}$ is continuous in ${{{\widetilde {\bf{f}}}_k}}$ and $\textrm{li}{\textrm{m}_{j \to \infty }}{\bf{f}}_k^{i_j} = {\bf{f}}_k^*$, then there must exist, for any fixed ${{\bf{f}}_k} \in \mathbb{C}{_{\succ}}({\bf{f}}_k^*)$, an integer ${I_{{{\bf{f}}_k}}}$ such that
\begin{equation}
\overline {\bf{R}} ({{\bf{f}}_k},{\bf{f}}_k^{{i_j}},{{\bf{F}}_k})\succ{\bf{0}},\forall j > {I_{{{\bf{f}}_k}}}.
\end{equation}
This implies that there must exist a sufficiently large $T$ such that
\begin{equation}
\mathbb{C}{_{\succ}}({\bf{f}}_k^*) \subseteq \mathbb{C}{_{\succ}}({\bf{f}}_k^{{i_j}}) \subset \mathbb{C}{_{\succeq}}({\bf{f}}_k^{{i_j}}),\forall j > T.
\end{equation}

Since $\{ {\bf{f}}_k^{{i_j} + 1}\}  \in \mathbb{S}({\bf{f}}_k^{{i_j}})$, we can see that
\begin{equation} \label{prove_1}
P({{\bf{f}}_k}) \ge P({\bf{f}}_k^{{i_j} + 1}),\;\forall {{\bf{f}}_k} \in \mathbb{C}{_{\succ}}({\bf{f}}_k^*) \subset \mathbb{C}{_{\succeq}}({\bf{f}}_k^{{i_j}}).
\end{equation}
Moreover, since $P( \cdot )$ is a continuous function, we have by letting ${j \to \infty }$ in (\ref{prove_1}) that
\begin{equation}
P({{\bf{f}}_k}) \ge P({\bf{f}}_k^{{**}}),\;\forall {{\bf{f}}_k} \in \mathbb{C}{_{\succ}}({\bf{f}}_k^*).
\end{equation}
It follows from the continuity of ${\overline {\bf{R}} ({{\bf{f}}_k},{{\widetilde {\bf{f}}}_k},{{\bf{F}}_k})}$ that
\begin{equation} \label{prove_2}
P({{\bf{f}}_k}) \ge P({\bf{f}}_k^{{**}}),\;\forall {{\bf{f}}_k} \in \mathbb{C}{_{\succeq}}({\bf{f}}_k^*).
\end{equation}
One the other hand, we can infer from the second step that
\begin{equation} \label{prove_3}
P({{\bf{f}}_k^*}) = P({\bf{f}}_k^{{**}}).
\end{equation}
Furthermore, since ${\bf{f}}_k^{{i_j}}$ is feasible to problem (\ref{iterative_SDP3}) and $\overline {\bf{R}} ({\bf{f}}_k^{{i_j}},{\bf{f}}_k^{{i_j}},{{\bf{F}}_k}) = {\bf{R}}({\bf{f}}_k^{{i_j}})$, we have ${\bf{f}}_k^{{i_j}} \in \mathbb{C}{_{\succeq}}({\bf{f}}_k^{{i_j}}) $. It follows that ${\bf{f}}_k^{{*}} \in \mathbb{C}{_{\succeq}}({\bf{f}}_k^{{*}}) $. Combining this with (\ref{prove_2}) and (\ref{prove_3}), we obtain ${\bf{f}}_k^* \in \mathbb{S}({\bf{f}}_k^*)$.

According to \emph{Lemma} \ref{lamma1}, we have ${\bf{f}}_k^*{\bf{f}}_k^{*H} = {\bf{F}}_k^*$. Then we argue that $\{ {\bf{f}}_k^*,{\bf{F}}_k^*,{\rho _k^*}\} $ satisfy the KKT conditions of problem (\ref{iterative_SDP3}).
With \emph{Lemma} \ref{lamma2} and ${\bf{f}}_k^* \in \mathbb{S}({\bf{f}}_k^*)$, there must exist optimal Lagrange multipliers as follows (fix ${\alpha _k},{\beta _k}$)
\begin{equation}
\begin{array}{l}
\widetilde {\bf{U}}_k^* \buildrel \Delta \over = \left[ {\begin{array}{*{20}{c}}
{\overline {\bf{U}} _k^*}&{{\bf{u}}_k^*}\\
{{\bf{u}}_k^{*H}}&{u_k^*}
\end{array}} \right],\;\widetilde {\bf{V}}_{kj}^* \buildrel \Delta \over = \left[ {\begin{array}{*{20}{c}}
{\overline {\bf{V}} _{kj}^*}&{{\bf{v}}_{kj}^*}\\
{{\bf{v}}_{kj}^{*H}}&{v_{kj}^*}
\end{array}} \right],\\
\widetilde {\bf{X}}_k^* \buildrel \Delta \over = \left[ {\begin{array}{*{20}{c}}
{\overline {\bf{X}} _k^*}&{{\bf{x}}_k^*}\\
{{\bf{x}}_k^{*H}}&{x_k^*}
\end{array}} \right],\;\widetilde {\bf{Y}}_{kj}^* \buildrel \Delta \over = \left[ {\begin{array}{*{20}{c}}
{\overline {\bf{Y}} _{kj}^*}&{{\bf{y}}_{kj}^*}\\
{{\bf{y}}_{kj}^{*H}}&{y_{kj}^*}
\end{array}} \right],\\
{\bf{Z}}_k^*,\;\;a_{kj}^*,\;\;b_{kj}^*,\;\;c_{kj}^*,\;{\mkern 1mu} d_{kj}^*.
\end{array}
\end{equation}
Together with $\{{{{\bf{F}}_k^*},{{\bf{f}}_k^*},{\alpha _k^*},{\beta _k^*},{\lambda _{kj}^*},{\mu _{kj}^*},{p_{kj}^*},{q_{kj}^*}}\}$, we can see that the following KKT conditions of problem $\mathcal{P}({{\bf{f}}_k^*})$ hold
\begin{equation} \label{primal feasibility}
\begin{array}{l}
{{\bf{U}}_k}( {{\bf{F}}_k^*,{{\{ {p_{kj}^*} \}}_{j \ne k}},\lambda _{kk}^*,\alpha _k^*} )\succeq{\bf{0}},\\
{{\bf{V}}_{kj}}( {{\bf{F}}_j^*,p_{kj}^*,\lambda _{kk}^*} )\succeq{\bf{0}},\;j \ne k,\\
{{\bf{X}}_k}( {{\bf{F}}_k^*,{{\{ {q_{kj}^*} \}}_{j \ne k}},\mu _{kk}^*,\beta _k^*} )\succeq{\bf{0}},\\
{{\bf{Y}}_{kj}}( {{\bf{F}}_j^*,q_{kj}^*,\mu _{kj}^*} )\succeq{\bf{0}},\;j \ne k,\\
{\bf{F}}_k^* \succeq{\bf{0}},\; {\lambda _{kj}^*}\geq 0,\; {\mu _{kj}^*}\geq 0,\; \forall k \in \mathcal{K},
\end{array}
\end{equation}
\begin{equation}\label{dual feasibility}
\begin{array}{l}
{{\bf{Z}}_k^*} \buildrel \Delta \over = \\
{\bf{I}} - \frac{1}{{{\gamma _k}}}( {{{\overline {\bf{U}} }_k^*} + {{\bf{u}}_k^*}\widehat {\bf{h}}_{kk}^H + \widehat {\bf{h}}_{kk}^{}{\bf{u}}_k^{*H} + {u_k^*}\widehat {\bf{h}}_{kk}^{}\widehat {\bf{h}}_{kk}^H} )+\\
\sum\limits_{j \ne k}^K {( {{{\overline {\bf{V}} }_{jk}^*} + {{\bf{v}}_{jk}^*}\widehat {\bf{h}}_{jk}^H + \widehat {\bf{h}}_{jk}^{}{\bf{v}}_{jk}^{*H} + {v_{jk}^*}\widehat {\bf{h}}_{jk}^{}\widehat {\bf{h}}_{jk}^H} )} -\\
( {{{\overline {\bf{X}} }_k^*} + {{\bf{x}}_k^*}\widehat {\bf{h}}_{kk}^H + \widehat {\bf{h}}_{kk}^{}{\bf{x}}_k^{*H} + {x_k}\widehat {\bf{h}}_{kk}^{}\widehat {\bf{h}}_{kk}^H} ) -\\
\sum\limits_{j \ne k}^K {( {{{\overline {\bf{Y}} }_{jk}^*} + {{\bf{y}}_{jk}^*}\widehat {\bf{h}}_{jk}^H + \widehat {\bf{h}}_{jk}^{}{\bf{y}}_{jk}^{*H} + {y_{jk}^*}\widehat {\bf{h}}_{jk}^{}\widehat {\bf{h}}_{jk}^H} )} \succeq {\bf{0}},\\
{\widetilde {\bf{U}}_k^*} \succeq {\bf{0}},\; \widetilde {{\bf{V}}}_k^*\succeq {\bf{0}} ,\;{\widetilde {\bf{X}}_k^*} \succeq {\bf{0}},\;{\widetilde {\bf{Y}}_k^*} \succeq {\bf{0}},\;\\
{a_{kk}^*} \buildrel \Delta \over = {u_k^*}\eta _{kk}^2 - \textrm{Tr}( {{{\overline {\bf{U}} }_k^*}} ) \ge 0,\\
{a_{kj}^*} \buildrel \Delta \over = {v_{kj}^*}\eta _{kj}^2 - \textrm{Tr}( {{{\overline {\bf{V}} }_{kj}^*}} ) \ge 0,\;j \ne k,\\
{b_{kk}^*} \buildrel \Delta \over = {x_k^*}\eta _{kk}^2 - \textrm{Tr}( {{{\overline {\bf{X}} }_k^*}} ) \ge 0,\\
{b_{kj}^*} \buildrel \Delta \over = {y_{kj}^*}\eta _{kj}^2 - \textrm{Tr}( {{{\overline {\bf{Y}} }_{kj}^*}} ) \ge 0,\;j \ne k,\\
{c_{kj}^*} \buildrel \Delta \over = {u_k}^* - {v_{kj}^*} \ge 0,\; {d_{kj}^*} \buildrel \Delta \over = {y_{kj}^*} - {x_{k}^*} \ge 0,
\end{array}
\end{equation}
\begin{equation} \label{complementarity}
\begin{array}{l}
\textrm{Tr}( {\widetilde {\bf{U}}_k^*{{\bf{U}}_k}( {{\bf{F}}_k^*,{{\{ {p_{kj}^*} \}}_{j \ne k}},\lambda _{kk}^*,\alpha _k^*} )} ) = 0,\\
\textrm{Tr}( {\widetilde {\bf{V}}_{kj}^*{{\bf{V}}_{kj}}( {{\bf{F}}_j^*,p_{kj}^*,\lambda _{kk}^*} )} ) = 0,\;j \ne k,\\
\textrm{Tr}( {\widetilde {\bf{X}}_k^*{{\bf{X}}_k}( {{\bf{F}}_k^*,{{\{ {q_{kj}^*} \}}_{j \ne k}},\mu _{kk}^*,\beta _k^*} )} ) = 0,\\
\textrm{Tr}( {\widetilde {\bf{Y}}_{kj}^*{{\bf{Y}}_{kj}}( {{\bf{F}}_j^*,q_{kj}^*,\mu _{kj}^*} )} ) = 0,\;j \ne k,\\
\textrm{Tr}( {\widetilde {\bf{Z}}_k^*{\bf{F}}_k^*} ) = 0,\\
a_{kj}^*\lambda _{kj}^* = 0,\;b_{kj}^*\mu _{kj}^* = 0,\;\\
c_{kj}^*p_{kj}^* = 0,\;d_{kj}^*q_{kj}^* = 0,\;j \ne k,\;\forall j,k \in \mathcal{K},
\end{array}
\end{equation}
where we have used the fact ${\bf{f}}_k^*{\bf{f}}_k^{*H} = {\bf{F}}_k^*$. Note that (\ref{primal feasibility}) denotes the primal feasible conditions, (\ref{dual feasibility}) denotes the first-order necessary optimality conditions and dual feasibility conditions, (\ref{complementarity}) denotes the complementarity conditions. Eqs. (\ref{primal feasibility})-(\ref{complementarity}) imply that $\{ {\bf{f}}_k^*,{\bf{F}}_k^*,{\rho _k^*}\} $ is a KKT point of problem (\ref{iterative_SDP3}). From the above analysis, we conclude that $\{ {\bf{f}}_k^*,{\bf{F}}_k^*,{\rho _k^*}\} $ is a KKT point of problem (\ref{SDRproblem}). This completes the proof.

\end{appendices}


\begin{thebibliography}{1}

\bibitem{varshney2008transporting}
L.~R. Varshney, ``Transporting information and energy simultaneously,'' in \emph{IEEE Int. Symp. Inf. Theory (ISIT)}, pp. 1612--1616, Jul. 2008.

\bibitem{grover2010shannon}
P.~Grover and A.~Sahai, ``Shannon meets Tesla: Wireless information and power transfer.'' in \emph{IEEE Int. Symp. Inf. Theory (ISIT)}, pp. 2363--2367, Jun. 2010.

\bibitem{OntheTransferofInformation}
A.~Fouladgar and O.~Simeone, ``On the transfer of information and energy in multi-user systems,'' \emph{IEEE Commun. Lett.}, vol.~16, no.~11, pp. 1733--1736, Nov. 2012.

\bibitem{Twowaycommunication}
P.~Popovski, and O.~Simeone, ``Two-way communication with energy exchange,'' in \emph{Proc. IEEE Inf. Theory Workshop}, pp. 592--596, Sept. 2012.

\bibitem{MIMOBroadcastingfor}
R.~Zhang and C.~K. Ho, ``MIMO broadcasting for simultaneous wireless information and power transfer,'' \emph{IEEE Trans. Wireless Commun.}, vol.~12, no.~5, pp. 1989--2001, May 2013.

\bibitem{RobustBeamformingforWireless}
Z.~Xiang and M.~Tao, ``Robust beamforming for wireless information and power transmission,'' \emph{IEEE Wireless Commun. Letters}, vol.~1, no.~4, pp. 372--375, Aug. 2012.

\bibitem{MultiuserMISOBeamforming}
J.~Xu, L.~Liu, and R.~Zhang, ``Multiuser MISO beamforming for simultaneous wireless information and power transfer,'' \emph{IEEE Trans. Signal Process.}, vol.~62, no.~18, pp. 4798--4810, Sept. 2014.

\bibitem{ADynamicPowerSplittingApproach}
L.~Liu, R.~Zhang, and K.-C. Chua, ``Wireless information and power transfer: A dynamic power splitting approach,'' \emph{IEEE Trans. Commun.}, vol.~61, no.~9, pp. 3990--4001, Sept. 2013.

\bibitem{ArchitectureDesign}
X.~Zhou, R.~Zhang, and C.~K. Ho, ``Wireless information and power transfer: Architecture design and rate-energy tradeoff,'' \emph{IEEE Trans. Commun.}, vol.~61, no.~11, pp. 4754--4767, Nov. 2013.

\bibitem{EnergyharvestinginanOSTBC}
B.~Chalise, Y.~Zhang, and M.~Amin, ``Energy harvesting in an OSTBC based amplify-and-forward MIMO relay system,'' in \emph{Proc. IEEE Int. Conf. on Acoustis, Speech and Signal Processing}, pp. 3201--3204, Mar. 2012.

\bibitem{Li2014Beamforming}
Q.~Li, Q.~Zhang, and J.~Qin, ``Beamforming in non-regenerative two-way multi-antenna relay networks for simultaneous wireless information and power transfer,'' \emph{IEEE Trans. Wireless Commun.}, vol.~13, no.~10, pp. 5509--5520, Oct. 2014.

\bibitem{Chen2015Distributed}
H.~Chen, Y.~Li, Y.~Jiang, Y.~Ma, and B.~Vucetic, ``Distributed power splitting for SWIPT in relay interference channels using game theory,'' \emph{IEEE Trans. Wireless Commun.}, vol.~14, no.~1, pp. 410--420, Jan. 2015.

\bibitem{JointTransmitBeamforming}
Q. Shi, W. Xu, L. Liu and R. Zhang, ``Joint transmit beamforming and receive power splitting for MISO swipt systems,'' \emph{IEEE Trans. Wireless Commun.}, vol.~13, no.~6, pp. 3269--3280, Jun. 2014.

\bibitem{Two-UserMIMOInterferenceChannel}
J.~Park and B.~Clerckx, ``Joint wireless information and energy transfer in a two-user MIMO interference channel,'' \emph{IEEE Trans. Wireless Commun.}, vol.~12, no.~8, pp. 4210--4221, Aug. 2013.

\bibitem{K-UserMIMOInterferenceChannel}
------, ``Joint wireless information and energy transfer in a $K$-user MIMO interference channel,'' \emph{IEEE Trans. Wireless Commun.}, vol.~13, no.~10, pp. 5781--5796, Oct. 2014.

\bibitem{BeamformingforMISO}
S. Timotheou, I. Krikidis, G. Zheng and B. Ottersten, ``Beamforming for MISO interference channels with QoS and RF energy transfer,'' \emph{IEEE Trans. Wireless Commun.}, vol.~13, no.~5, pp. 2646--2658, May 2014.

\bibitem{JointBeamformingAnd}
Q.~Shi, W.~Xu, T.-H. Chang, Y.~Wang, and E.~Song, ``Joint beamforming and power splitting for MISO interference channel with SWIPT: An SOCP relaxation and decentralized algorithm,'' \emph{IEEE Trans. Signal Process.}, vol.~62, no.~23, pp. 6194--6208, Dec. 2014.

\bibitem{Mguyen2015Opportunistic}
V.-D. Mguyen, Son Dinh-Van, and O. S. Shin, ``Opportunistic relaying with wireless energy harvesting in a cognitive radio system,'' \emph{Proc. IEEE WCNC}, Mar. 2015.

\bibitem{Relayprecoder}
P.~Ubaidulla and A.~Chockalingam, ``Relay precoder optimization in MIMO-relay networks with imperfect CSI,'' \emph{IEEE Trans. Signal Process.}, vol.~59, no.~11, pp. 5473--5484, Nov. 2011.

\bibitem{Robustdownlinkpower}
M. Biguesh, S. Shahbazpanahi, and A. B. Gershman, ``Robust downlink power control in wireless cellular systems,'' \emph{EURASIP J. Wireless Commun. Netw.}, vol.~2, pp. 261--272, 2004.

\bibitem{RobustPowerallocation}
M. Payaro, A. Pascual-Iserte, and M. A. Lagunas, ``Robust power allocation designs for multiuser and multiantenna downlink communication systems through convex optimization,'' \emph{IEEE J. Sel. Areas Commun.}, vol.~25, no.~7, pp. 1390--1401, Sept. 2007.

\bibitem{ConvexOptimization}
S.~Boyd and L.~Vandenberghe, \emph{Convex Optimization}.\hskip 1em plus 0.5em minus 0.4em\relax Cambridge, U.K.: Cambridge Univ. Press, 2004.

\bibitem{boyd1994linear}
S.~P. Boyd, L.~El~Ghaoui, E.~Feron, and V.~Balakrishnan, \emph{Linear Matrix Inequalities in System and Control Theory}.\hskip 1em plus 0.5em minus 0.4em\relax SIAM, vol.~15, 1994.

\bibitem{Distributedrobustmulti}
C.~Shen, T.-H. Chang, K.-Y. Wang, Z.~Qiu, and C.-Y. Chi, ``Distributed robust multicell coordinated beamforming with imperfect CSI: An ADMM approach,'' \emph{IEEE Trans. Signal Process.}, vol.~60, no.~6, pp. 2988--3003, Jun. 2012.

\bibitem{OptimizationofMIMOrelays}
B.~Chalise and L.~Vandendorpe, ``Optimization of MIMO relays for multipoint-to-multipoint communications: Nonrobust and robust designs,'' \emph{IEEE Trans. Signal Process.}, vol.~58, no.~12, pp. 6355--6368, Dec. 2010.

\bibitem{Semidefiniterelaxation}
Z.-Q. Luo, W.-K. Ma, A.-C. So, Y.~Ye, and S.~Zhang, ``Semidefinite relaxation of quadratic optimization problems,'' \emph{IEEE Trans. Signal Process. Mag.}, vol.~27, no.~3, pp. 20--34, May 2010.


\bibitem{UsingSeDuMi}
J.~F. Sturm, ``Using sedumi 1.02, a matlab tool for optimization over symmetric cones,'' \emph{Optim. Methods Softw.}, vol. 11-12, pp. 625--653, 1999.

\bibitem{Solvingsemidefinite}
R. H. T\"ut\"unc\"u, K. C. Toh and M. J. Todd, ``Solving semidefinite-quadratic-linear programs using SDPT3,'' \emph{Math. Program., Ser. B}, vol.~95, no.~2, pp. 189--217, 2003.

\bibitem{AdaptiveFilterTheory}
S.~Haykin, \emph{Adaptive Filter Theory}.\hskip 1em plus 0.5em minus 0.4em\relax 4th ed. Englewood Cliffs, NJ: Prentice-Hall, 2002.

\bibitem{Applicationsofsecond}
M. S. Lobo, L. Vandenberghe, S. Boyd, and H. Lebret, \emph{Applications of second order cone programming}.\hskip 1em plus 0.5em minus 0.4em\relax Linear Algebra APP., 1998.


\bibitem{yuille2003concave}
A.~L. Yuille and A.~Rangarajan, ``The concave-convex procedure,'' \emph{Neural Computation}, vol.~15, no.~4, pp. 915--936, 2003.

\bibitem{lanckriet2009convergence}
G.~R. Lanckriet and B.~K. Sriperumbudur, ``On the convergence of the concave-convex procedure,'' in \emph{Advances in Neural Information Processing Systems}, pp. 1759--1767, 2009.

\bibitem{cheng2012joint}
Y.~Cheng and M.~Pesavento, ``Joint optimization of source power allocation and distributed relay beamforming in multiuser peer-to-peer relay networks,'' \emph{IEEE Trans. Signal Process.}, vol.~60, no.~6, pp. 2962--2973, Jun. 2012.

\bibitem{Wang2014Outage}
K.-Y. {Wang}, A.~{Man-Cho So}, T.-H. {Chang}, W.-K. {Ma}, and C.-Y. {Chi}, ``{Outage constrained robust transmit optimization for multiuser MISO downlinks: tractable approximations by conic optimization},'' \emph{IEEE Trans. Signal Process.}, vol.~62, no.~21, pp. 5690--5705, Nov. 2014.

\bibitem{CVX}
C.~R. Inc., ``Cvx: Matlab software for disciplined convex programming,'' Available: http://cvxr.com/cvx, Sept. 2012.

\bibitem{2014arXiv1403.3196S}
Q.~{Shi}, W.~{Xu}, J.~{Wu}, E.~{Song}, and Y.~{Wang}, ``{Secure beamforming for MIMO broadcasting with wireless information and power transfer},'' available on-line at arXiv : 1403.3196, Mar. 2014.



%
%
\end{thebibliography}

\end{document}